\documentclass[iop]{emulateapj}
\usepackage{graphicx}
\usepackage{natbib}
\usepackage{amsmath}
\newcommand{\degree}{\hbox{$^\circ$}}

\newcommand{\ltsimeq}{\la}
\newcommand{\gtsimeq}{\ga}

\newcommand{\msun}{M$_{\odot}$}

\newcommand{\HI}{H{\sc i}}
\newcommand{\HII}{H{\sc ii}}

\accepted{ApJ July 2015}
\shortauthors{McQuinn et al.}
\shorttitle{Leo~P: An Unquenched Very Low-Mass Galaxy}

\begin{document}
\title{Leo~P: An Unquenched Very Low-Mass Galaxy\protect\footnotemark[*]}
\footnotetext[*]{Based on observations made with the NASA/ESA Hubble Space Telescope, obtained from the Data Archive at the Space Telescope Science Institute, which is operated by the Association of Universities for Research in Astronomy, Inc., under NASA contract NAS 5-26555.}
\author{Kristen~B.~W. McQuinn\altaffilmark{1,2}, 
Evan D.~Skillman\altaffilmark{1},
Andrew Dolphin\altaffilmark{3},
John M.~Cannon\altaffilmark{4},
John J.~Salzer\altaffilmark{5},
Katherine L.~Rhode\altaffilmark{5},
Elizabeth A.~K.~Adams\altaffilmark{6},
Danielle Berg\altaffilmark{1,7},
Riccardo Giovanelli\altaffilmark{8},
L\'{e}o Girardi\altaffilmark{9},
Martha P.~Haynes\altaffilmark{8}
}

\altaffiltext{1}{Minnesota Institute for Astrophysics, School of Physics and
Astronomy, 116 Church Street, S.E., University of Minnesota,
Minneapolis, MN 55455; \ {\it kmcquinn@astro.umn.edu}}
\altaffiltext{2}{University of Texas at Austin, McDonald Observatory, 2515 Speedway, Stop C1402, Austin, Texas 78712}
\altaffiltext{3}{Raytheon Company, 1151 E. Hermans Road, Tucson, AZ 85756}
\altaffiltext{4}{Department of Physics and Astronomy, 
Macalester College, 1600 Grand Avenue, Saint Paul, MN 55105}
\altaffiltext{5}{Department of Astronomy, Indiana University, 727 East 3rd Street, Bloomington, IN 47405}
\altaffiltext{6}{ASTRON, the Netherlands Institute for Radio Astronomy, Postbus 2, 7990 AA, Dwingeloo, The Netherlands}
\altaffiltext{7}{Center for Gravitation, Cosmology and Astrophysics, Department of Physics, University of Wisconsin Milwaukee, 1900 East Kenwood Boulevard, Milwaukee, WI 53211}
\altaffiltext{8}{Center for Radiophysics and Space Research, Space Sciences Building, Cornell University, Ithaca, NY 14853}
\altaffiltext{9}{Osservatorio Astronomico di Padova, INAF, Vicolo dell'Osservatorio 5, I-35122 Padova, Italy}

\begin{abstract}
Leo~P is a low-luminosity dwarf galaxy discovered through the blind \HI\ Arecibo Legacy Fast ALFA (ALFALFA) survey. The \HI\ and follow-up optical observations have shown that Leo~P is a gas-rich dwarf galaxy with active star formation, an underlying older population, and an extremely low oxygen abundance. We have obtained optical imaging with the \textit{Hubble Space Telescope} to two magnitudes below the red clump in order to study the evolution of Leo~P. We refine the distance measurement to Leo~P to be $1.62\pm0.15$ Mpc, based on the luminosity of the horizontal branch stars and 10 newly identified RR~Lyrae candidates. This places the galaxy at the edge of the Local Group, $\sim0.4$ Mpc from Sextans~B, the nearest galaxy in the NGC~3109 association of dwarf galaxies of which Leo~P is clearly a member. The star responsible for ionizing the \HII\ region is most likely an O7V or O8V spectral type, with a stellar mass $\gtsimeq25$ \msun. The presence of this star provides observational evidence that massive stars at the upper-end of the initial mass function are capable of being formed at star formation rates as low as $\sim10^{-5}$ \msun\ yr$^{-1}$. The best-fitting star formation history derived from the resolved stellar populations of Leo~P using the latest PARSEC models shows a relatively constant star formation rate over the lifetime of the galaxy. The modeled luminosity characteristics of Leo~P at early times are consistent with low-luminosity dSph Milky Way satellites, suggesting that Leo~P is what a low-mass dSph would look like if it evolved in isolation and retained its gas. Despite the very low mass of Leo~P, the imprint of reionization on its star formation history is subtle at best, and consistent with being totally negligible.  The isolation of Leo~P, and the total quenching of star formation of Milky Way satellites of similar mass, implies that local environment dominates the quenching of the Milky Way satellites.
\end{abstract} 


\section{Introduction\label{intro}}
\subsection{Evolution of Very Low-Mass Galaxies}
Galaxies at the faint-end of the luminosity function (LF) have provided vital tests for our understanding of structure formation and evolution. The dearth of both observable low-mass dwarf galaxies \citep[``missing satellite problem''; e.g.,][]{Kauffmann1993, Klypin1999, Moore1999} and of higher surface brightness dwarf galaxies \citep[``too big to fail'';][]{Boylan-Kolchin2011} compared to simulations based on the $\Lambda$CDM model has proven to be a multi-decade challenge. These issues have been partly alleviated with higher resolution simulations that include more baryonic effects \citep[e.g.,][]{Brooks2013, Benitez-Llambay2014, Sawala2015, Onorbe2015}. In addition, large sky surveys have detected an increasing number of satellite systems, providing progressively better agreement between theory and observations for the missing satellite problem \citep[e.g.,][]{Koposov2015}. While these results have reduced the inconsistencies, future revisions to structure formation theories will undoubtedly be anchored by observational constraints from studies of very low-mass systems. 

Furthermore, galaxies at the faint-end of the LF are broadening our understanding of star formation and galaxy evolution. The star formation properties of gas-rich, low-mass galaxies encompass a larger range of parameters for a given stellar mass than more massive, ``main-sequence'' galaxies \citep{Brinchmann2004, Bothwell2009, Huang2012, McQuinn2015a}. It is unclear whether this is due to an inherently stochastic star formation process at low molecular gas column densities and metallicities, or to a variation in the primary processes that regulate star formation. 

\begin{figure*}[ht]
\includegraphics[width=\textwidth]{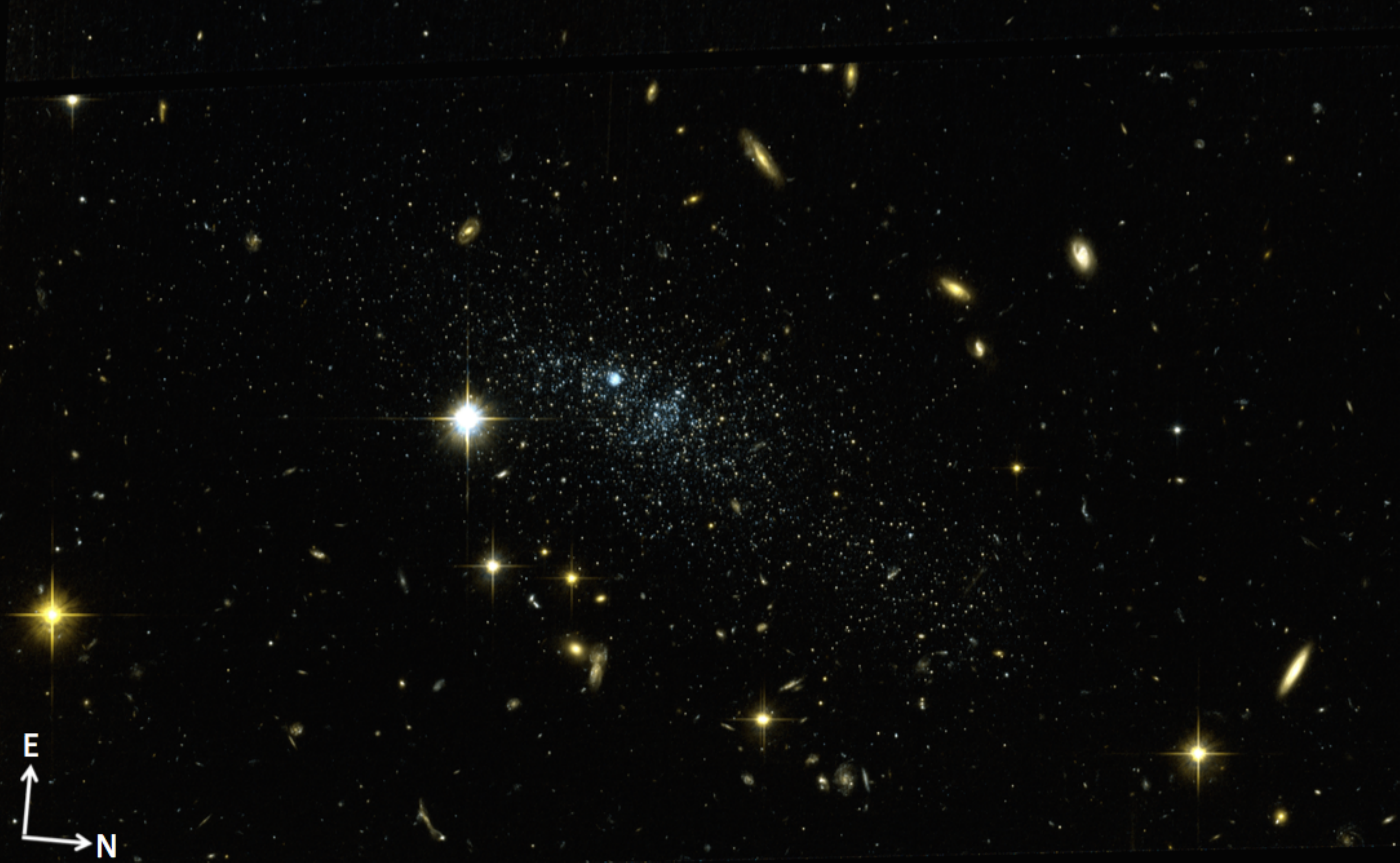}
\caption{Three-color HST image of Leo~P showing both the main star forming complex and underlying older stellar population extending to greater radii. The image was created by combining the F475W band image (Blue), the average of F475W and F814W band images (Green), and the F814W band image (Red). The field of view shown here is approximately 3.1\arcmin$\times$1.8\arcmin. Note the dither pattern used did not cover the chip gap which is visible across the top of the image.}
\label{fig:image}
\end{figure*}

In the nearby Universe, much work has been done to understand the evolutionary history of the rich population of nearby low-mass galaxies \citep[e.g.,][among others]{Skillman1989, Hunter2004, Begum2008, Lee2009b, Dalcanton2009, McQuinn2010a, Weisz2011, Cannon2011c, McConnachie2012, Hunter2012, Lelli2014a, Weisz2014a, Boyer2015a, Boyer2015b, McQuinn2015a,McQuinn2015c}. The interpretations are often complicated because relative differences between low-mass galaxies can depend significantly on large-scale environmental differences. This latter factor is often described as the morphology-density relation, or the prevalence of gas-poor dwarfs (i.e., dwarf spheroidals or dSphs) to be located in higher galactic density environments whereas gas-rich dwarfs (i.e., dwarf irregulars or dIrrs) are predominantly found in lower density field environments \citep[e.g.,][]{Einasto1974, Binggeli1987, VandenBergh1994, Cote2009}. The closest (and most accessible) low-mass galaxies are coevolving with the Milky Way, and thus disentangling the impact of the internal versus external evolutionary processes is a significant challenge. Ideally, one could study the properties of a low-mass system located $\textit{outside}$ of a group environment. While such low-mass, low-luminosity systems are expected to be numerous, they have generally eluded detection outside of the immediate vicinity in the Local Group (LG) due to their faint and sometimes low surface brightness nature.

An exception is Leo~P, a low-luminosity ($M_V = -9.27$ mag), gas-rich galaxy discovered just outside the zero-velocity boundary of the LG. Leo~P's combined properties of extremely low stellar mass ($5.6\times10^5$ \msun) and metallicity (3\% solar), significant gas content ($8.1\times10^5$ \msun), proximity (1.62 Mpc), and isolation (0.4 Mpc from its nearest neighbor and outside of any group environment) present a unique test for our understanding of both structure formation and galaxy evolution. Comparable in luminosity to some of the dSph satellites of the Milky Way and a few of the dIrrs inside the LG, Leo~P provides an opportune target for studying evolution at the faint end of the LF in a {\it simplified} system. Not only are many complex variables eliminated (e.g., tidal and ram pressure stripping, significant possible merger events, etc. that are present in the majority of galaxies detected in this luminosity regime), but Leo~P crosses over into the mass regime where (i) theories of structure formation predict difficulty in galaxies retaining their baryons due to stellar feedback processes \citep[e.g.,][]{Larson1974, Dekel1986, MacLow1999, Ferrara2000} and (ii) theories of the impact of reionization predict the photo-evaporization of the gas content and a sudden and irrevocable quenching of star formation \citep[e.g.,][]{Babul1992, Efstathiou1992, Thoul1996}.

In this work, we present a detailed study of the star formation history (SFH) and evolution of Leo~P using $HST$ optical imaging of Leo~P that reaches $\sim$2 mag below the red clump (RC). In Section~2 we discuss the observations, data reduction, and variable star identification. In Section~3, we refine the distance measurement to Leo~P using analysis of horizontal branch (HB) stellar populations and variable stars. In Section~4, we describe the methodology for deriving the best-fitting SFH, and in Section~5 we present measurements of the SFH and chemical evolution of Leo~P. In Section~6 we present a comparison of the evolution of Leo~P with previously studied Milky Way dSph galaxy satellites and three LG dIrrs. Our conclusions are summarized in Section~7. 

\subsection{A Brief Synopsis of Leo~P\label{story}}
Leo~P \citep[AGC~208583;][]{Giovanelli2013} was discovered through the Arecibo Legacy Fast ALFA survey \citep[ALFALFA;][]{Giovanelli2005, Haynes2011}. The ALFALFA survey is a blind extragalactic survey in the \HI\ 21 cm line covering over 7000 square degrees of high Galactic latitude sky. Follow-up optical observations at the location of the ALFALFA \HI\ detection, including WIYN 3.5m BVR and KPNO 2.1m H$\alpha$ imaging, confirmed the presence of both a resolved stellar population and a single \HII\ region \citep{Rhode2013}. 

Optical spectroscopy of the \HII\ region enabled a direct measurement of the auroral [O~III] $\lambda$4363 line, yielding an oxygen abundance of 12 $+$ log(O/H) $=7.17\pm0.04$ \citep{Skillman2013}, $\sim3$\% Z$_{\odot}$ \citep[based on a solar abundance of 12 $+$ log(O/H) $=8.68$;][]{Asplund2009}. This abundance measurement showed that Leo~P is the lowest metallicity, gas-rich galaxy known in the Local Volume. Its properties are consistent with the luminosity-metallicity relationship in \citet{Berg2012}.

Deeper optical imaging of the resolved stellar populations from the Large Binocular Telescope (LBT) enabled a distance measurement based on the tip of the red giant branch (TRGB) detection of $1.72^{+0.14}_{-0.40}$ \citep{McQuinn2013}. The large lower uncertainty is due to the sparseness of the red giant branch (RGB) stars in such a low-mass galaxy. This distance measurement placed Leo~P just outside the zero velocity boundary of the LG.

\HI\ spectral line imaging from the VLA shows a small amplitude rotation ($V_c=15\pm5$ km s$^{-1}$), with no obvious signs of interaction or in-falling gas at larger spatial scales \citep{Bernstein-Cooper2014}. These observations revealed that Leo~P has one of the lowest neutral gas mass of any known low-metallicity dwarf, with a mass ratio of gas to stars of $2:1$ and total mass to baryonic mass of $>15:1$. Follow-up high-sensitivity observations with CARMA of the CO (J$ = 1 \rightarrow 0$) transition did not detect any CO emission, but placed stringent upper limits on the CO luminosity of L$_{CO} < 2930$ K km s$^{-1}$ pc$^2$ (Warren et al.\ submitted). The molecular hydrogen mass remains uncertain as the ratio of CO to H$_2$ is not well constrained at low-metallicities. In Table~\ref{tab:properties} we summarize the basic properties measured in Leo~P from these previous studies.

\section{Data Processing\label{data}}
\subsection{Observations\label{obs}}
HST observations of Leo~P were obtained using the Advance Camera for Surveys (ACS) instrument \citep{Ford1998} in the F475W (Sloan g) and F814W (I) filters between April 23 and 26 2014 (GO 13376; PI McQuinn). The observations were taken over 17 orbits grouped in visits of $2-3$ orbits, and followed the ultra-deep field dither pattern between exposures, shifted by $2-3$ pixels to aid in the removal of hot pixels and to average-out the detector response. The observations were designed to reach a photometric depth of $\sim2$ mag below the RC to aid in (i) constraining the SFH at older times and (ii) measuring the distance from the HB stars and RR~Lyrae stars. 

Similar to the successful Local Cosmology from Isolated Dwarfs\protect\footnotemark[1]\footnotetext[1]{http://www.iac.es/project/LCID/} \citep[LCID, e.g.,][]{Skillman2014} program, our observation strategy was designed to optimally sample the light curves of short period variable stars \citep[e.g.,][]{Bernard2008, Bernard2009}, such as RR~Lyrae, which provide independent constraints on the distance to Leo~P and the SFH. To achieve this, each orbit was split between the two filters, with an alternating pattern for the exposures (i.e., F475W, F814W, F814W, F475W, etc.). In addition, the visits were executed sequentially and grouped within a 3-day time period, while minimizing orbits with a $\sim12$ hour cadence. These observations thus maximize sampling within the 17 orbits of a $\sim0.6$ day period RR~Lyrae star across 6 periods with minimal repetition of sampling at the same point in a light curve. The observational details are summarized in Table~\ref{tab:properties}.

HST Drizzlepac v2.0 was used to create mosaics in each filter. First, the charge transfer efficiency corrected images (i.e., {\tt flc} files) were processed with \textsc{Astrodrizzle} to remove cosmic-rays (CR), creating a set of CR cleaned images. Second, these CR cleaned images were processed through the task \textsc{tweakreg} to measure the astrometric shifts between the images. Third, the original {\tt flc.fits} images were processed with \textsc{Astrodrizzle} using the \textsc{tweakreg} shifts to create a combined mosaic in each filter. 

In Figure~\ref{fig:image}, we present a color image of Leo~P from the HST ACS observations created using the F475W and F814W mosaics, along with an averaged image of the two filters. As seen in Figure~\ref{fig:image}, the galaxy has an irregular stellar morphology that is well-resolved and a single \HII\ region clearly identifiable in the image.

\begin{table}
\begin{center}
\caption{Leo~P Properties and Observations}
\vspace{-10pt}
\label{tab:properties}
\end{center}
\begin{center}
\begin{tabular}{lrc}
\hline 
\hline 
						& 			& Updated Distance- \\
Parameter					& Value			& Dependent Values \\
\hline 
R.A. (J2000) 				& 10:21:45.1		& \nodata 			\\
Decl. (J2000)				& $+$18:05:17  	& \nodata			\\
Distance (Mpc)				& $1.72^{+0.14}_{-0.40}$ & $1.62\pm0.15$\\
$M_V$ (mag)				& $-9.41^{+0.17}_{-0.50}$& $-9.27\pm0.20$\\
$M_*$ (\msun)	from M/L		&  $5.7^{+0.4}_{-1.8}\times10^5$	& \nodata			\\
$M_*$ (\msun)	from SFH		& \nodata			& $5.6^{+0.4}_{-1.9}\times10^5$\\
M$_*$/L$_V$ 				& \nodata			& 1.25				\\
$M_{HI}$ (\msun)			& $9.3\times10^5$ 	& $8.1\times10^5$	\\
$V_{sys}$ (km s$^{-1}$)			& $260.8\pm2.5$			& \nodata	\\
$L_{CO}$ (K km s$^{-1}$ pc$^2$)& $<3300$		& $<2930$		\\
$12+$log(O/H)				& 7.17$\pm0.04$	& \nodata			\\
$L_{H\alpha}$ (erg s$^{-1}$)	& $6.2\times10^{36}$& $5.5\times10^{36}$	\\
SFR$_{H\alpha}$ (\msun\ yr$^{-1}$)& $4.9\times10^{-5}$& $4.3\times10^{-5}$\\
$A_V$ (mag)				& 0.071 			& \nodata			\\
Semi-major axis (\arcmin)		& 1.2				& \nodata			\\
Ellipticity					& 0.52			& \nodata			\\
Position angle (\degree)		& 335			& \nodata			\\
HST Program ID			& HST-GO-13376	& \nodata			\\
ACS Filters				& F475W; F814W	& \nodata			\\
Exposure F475W (s)			& 25,829			& \nodata			\\
Exposure F814W (s)			& 18,496			& \nodata			\\
\hline
\end{tabular}
\end{center}
\tablecomments{Summary of the properties of Leo~P based on measurements reported in \citet{Giovanelli2013, Rhode2013, McQuinn2013, Skillman2013, Bernstein-Cooper2014} and Warren et al.\ (submitted) and the revised distance measurement from this work (see Table~\ref{tab:distances}). Systemic velocity ($V_{sys}$) is in the Local Standard of Rest Kinematic frame. H$\alpha$-based SFR is based on the calibration by \citet{Kennicutt1994}. Foreground extinction estimate is based on the dust maps from \citet{Schlegel1998} with recalibration from \citet{Schlafly2011}.}
\end{table}

\subsection{Photometry and Color-Magnitude Diagram\label{cmd}}
Point spread function (PSF) photometry was performed on the flat, charge transfer efficiency corrected images ({\tt flc.fits}) with the {\tt DOLPHOT} photometry package, a modified version of {\tt HSTPHOT} with an ACS specific module  \citep{Dolphin2000}. The photometric catalog was filtered for well-recovered stars (i.e., output error flag $<8$) and with a signal-to-noise ratio $\geq5$. Point sources with high sharpness or crowding values were rejected (i.e., (V$_{sharp}+$I$_{sharp}$)$^2>0.075$; (V$_{crowd}+$I$_{crowd}$)$>0.8$). Sharpness indicates whether a point source is too broad (such as background galaxies) or too sharp (such as cosmic rays). Crowding measures how much brighter a star would be if nearby stars had not been fitted simultaneously; stars with higher values of crowding have higher photometric uncertainties. Spatial cuts were applied to the filtered photometric catalog. These cuts were based on the ellipticity (e $= $0.52), semi-major axis (1\arcmin.2), and position angle (PA $= 335\degree$) of Leo~P \citep{McQuinn2013}. The final elliptical parameters are listed in Table~\ref{tab:properties}. We expanded the spatial cuts for variable star identification (see Section~2.4), extending the search to include the halo of the galaxy.

Artificial star tests were performed to measure the completeness limits of the images using the same photometry package. Approximately 600k artificial stars were injected into the individual images following the spatial distribution of all sources in the non-filtered photometric catalog. The final artificial star lists were filtered using the same parameters that were used to produce the final photometry catalog.
 
\begin{figure}
\includegraphics[width=0.5\textwidth]{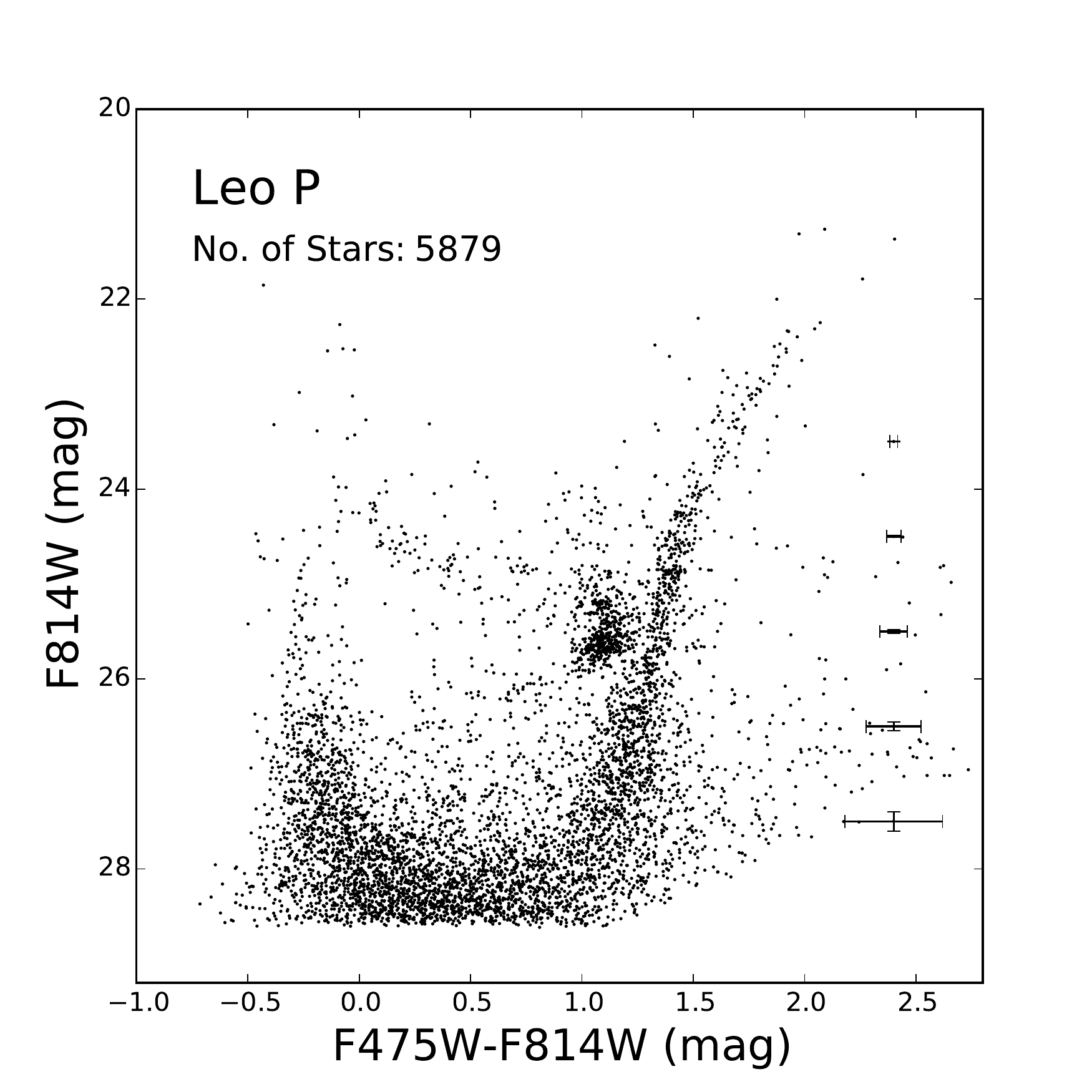}
\caption{The CMD of Leo~P from HST ACS imaging plotted to the 50\% completeness limit determined from artificial star tests. The photometry was corrected for Galactic extinction \citep{Schlafly2011}. Representative uncertainties per magnitude bin are plotted and include both photometric and completeness uncertainties. The MS, RGB, HB, and RC are seen in the CMD.}
\label{fig:cmd}
\end{figure}

Figure~\ref{fig:cmd} shows the CMD from the final photometry plotted to the 50\% completeness level as determined from the artificial star tests. Representative uncertainties per magnitude are plotted and include uncertainties from the photometry and uncertainties recovered from the artificial star tests. The photometry was corrected for extinction ($A_{F475W} = 0.086$ mag; $A_{F814W} = 0.039$ mag) based on the dust maps of \citet{Schlegel1998} with recalibration from \citet{Schlafly2011}. The main sequence (MS), RGB, HB, and the RC are identifiable in the CMD. In addition, there is a sparse population of red and blue helium burning stars (i.e., blue loop stars with F814W $\ltsimeq$ 25 mag). These stars are intermediate mass stars (2 \msun $\ltsimeq$ M $\ltsimeq $15 \msun) and are unambiguous signs of recent (t$\ltsimeq500$ Myr) star-formation activity \citep{Dohm-Palmer2002, McQuinn2011}. We investigate the star-forming properties in Leo~P in  Section~5. 

\begin{figure}[h]
\includegraphics[width=1.08\columnwidth]{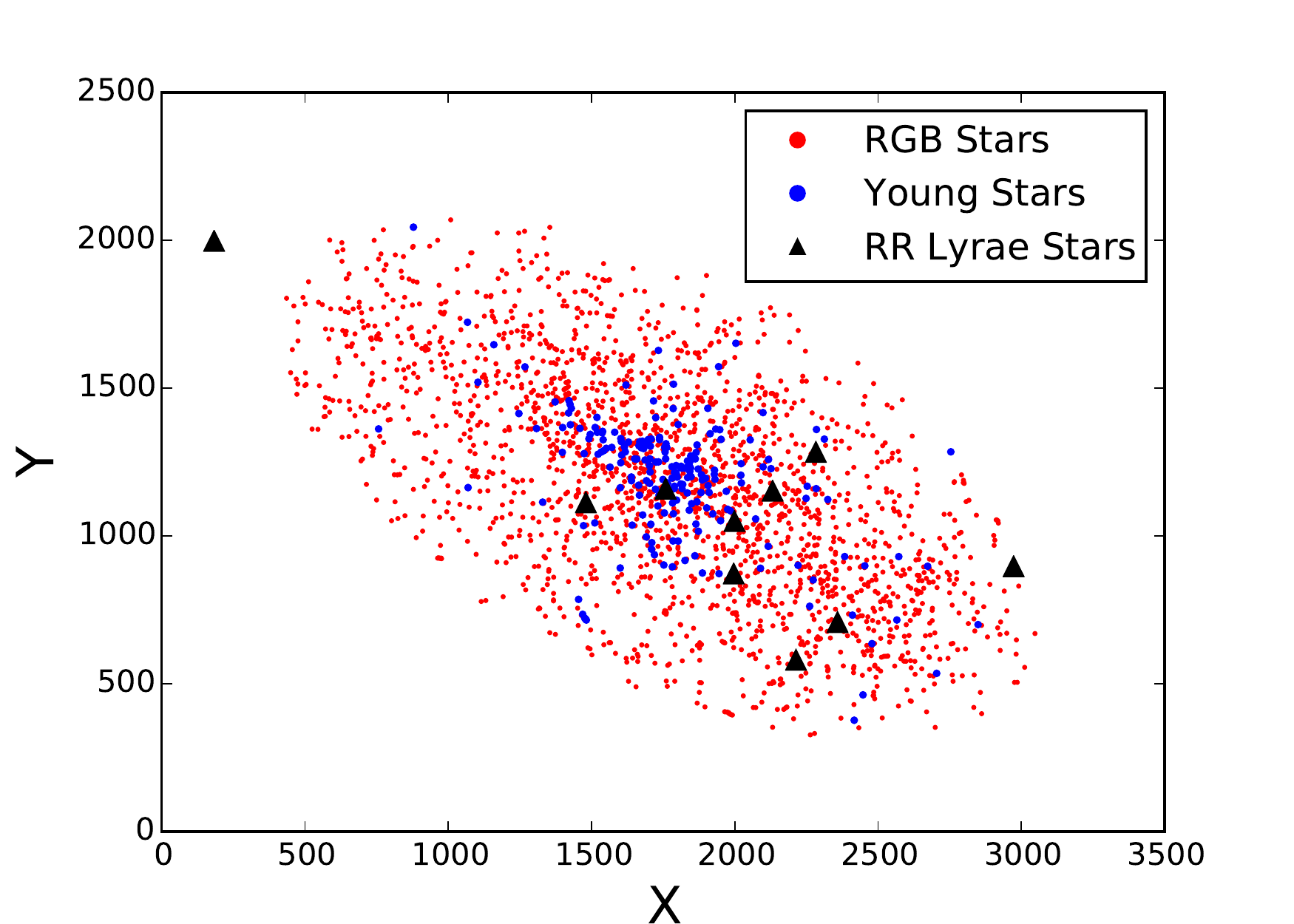}
\caption{Spatial distribution of ``young'' (age $\ltsimeq 300$ Myr) and ``old'' (age $\gtsimeq 300$ Myr) stars in Leo~P as defined in the text. The young stars are more centrally concentrated, with an irregular distribution. Despite the central concentration, there are still young stars located in the outer regions of Leo~P.}
\label{fig:spatial}
\end{figure}

Figure~\ref{fig:spatial} shows the spatial distribution of stars separated into course age bins of  ``young'' and ``old'' stars. Stars were classified as young and old based on their position in the CMD. Specifically, young stars were selected from the upper MS corresponding to a B or O spectral-type classification with lifetimes $\ltsimeq 300$ Myr (i.e., F475W $<$ 25.7 mag and F475W $-$ F814W $< 0.2$ mag), and the blue and red helium burning branch stars which have a similar maximum lifetime. In Figure~\ref{fig:cmd_highlights}, we overlay an isochrone of 300 Myr to demonstrate the location of young stars in the CMD. The old category is populated with RGB stars identified from the CMD. Despite the coarse time bins, the selection of young vs. old stars is robust. In Figure~\ref{fig:spatial}, the young stars are more centrally concentrated with an irregular distribution. A small number of young star candidates are located in the outer regions of the galaxy. The locations of the RR~Lyrae candidates, identified from the photometry, are also shown (see Section~2.4). 

\subsection{Resolved Stars in the \HII\ Region}
\begin{figure}[h]
\includegraphics[width=1.05\columnwidth]{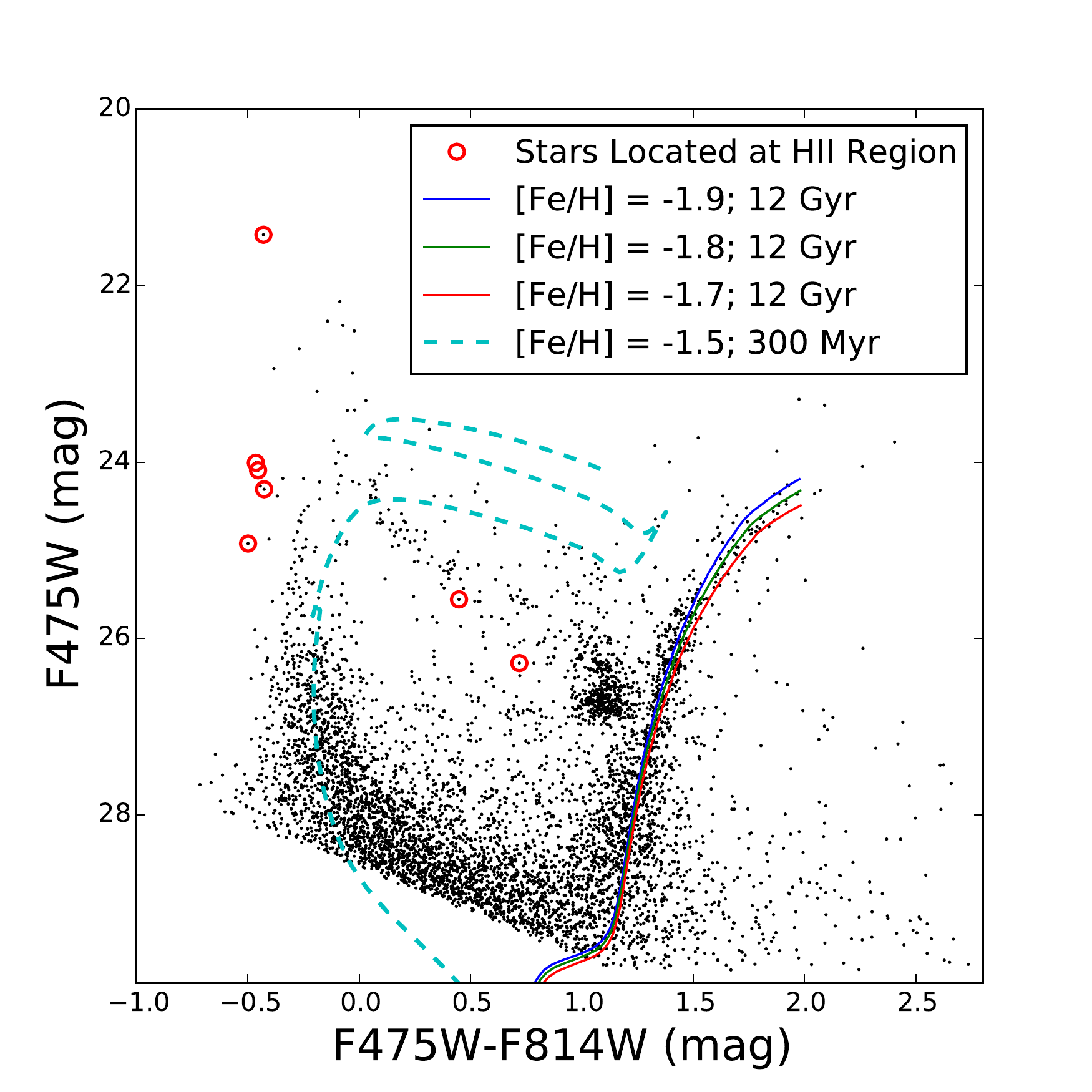}
\caption{CMD with the F475W plotted as the ordinate axis. Stars located at the position of the \HII\ region are circled in red. PARSEC stellar evolution isochrones are over plotted for a 12 Gyr population with metallicities varying between [Fe/H] of $-1.9$ to $-1.7$ and for a 300 Myr population with a metallicity of [Fe/H] $= -1.5$, as shown in the legend.}
\label{fig:cmd_highlights}
\end{figure}

The area within the one \HII\ region in the galaxy is resolved into 7 individual point sources in the $HST$ images. These sources were blended in previous, ground-based imaging of Leo~P, leading to some speculation on the nature of the star(s) responsible for ionizing the surrounding gas \citep[][see their \S4.1.2]{Rhode2013}. Figure~\ref{fig:cmd_highlights} shows the CMD of Leo~P with the F475W magnitude as the ordinate axis. Here, we highlight the stars located at the position of the \HII\ region in Leo~P in red, including the luminous star likely responsible for the ionization of the region. Note that these stars are bluer than expected in the CMD, which is most likely due to contamination by emission lines in the \HII\ region. The most luminous of these stars has de-reddened F475W and F814W mag of $21.509\pm0.001$ and $21.892\pm0.002$ respectively. Using the calibration of \citet{Saha2011} and an adopted distance modulus of $26.05\pm0.20$ (see Section~2.4 for discussion of the calibration and Section~3.3 for the adopted distance), the ACS filter luminosities transform to an absolute Johnson V mag of $-4.43$. A star of this magnitude in the upper main sequence region of the CMD is consistent with an O5V spectral type, according to the empirical catalog of absolute magnitudes of OB stars from \citet[][see their Table~1 of smoothed absolute magnitudes]{Wegner2000}. However, if this star is truly an O5V star, then the star with a similar color but $\sim1$ mag fainter at F475W $= 23.03$ mag would correspond to a star with an O9.5 spectral type, which would be hot enough to create a second \HII\ region. Since no \HII\ region is found around this fainter star, it is likely that both of these point sources are blended binary stars. High mass stars have been shown to have binary fractions in excess of 50\% \citep{Kobulnicky2014}. If we assume each of these MS stars is comprised of equal mass binaries, the brightest source located in the known \HII\ region could have two constituent O7 or O8 stars. The fainter source constituents would then be B2 or B3 stars, just below the limit for hosting a detectable \HII\ region.

\citet{Rhode2013} calculated a SFR based on the H$\alpha$ luminosity of the \HII\ region of $4.9\times10^{-5}$ \msun\ yr$^{-1}$. Using our adopted distance to Leo~P, this SFR is revised downward to a value of $4.3\times10^{-5}$ \msun\ yr$^{-1}$, compared with our SFR derived from the stellar populations over the past 4 Myr of $2.1\times10^{-5}$ \msun\ yr$^{-1}$ (see Section~5). In this regime, the H$\alpha$ luminosity has been shown to be a poor tracer of the actual star-formation activity due, in part, to the incomplete sampling of the high mass end of the initial mass function (IMF) \citep[e.g.,][]{Boselli2009, Goddard2010, Koda2012}. \citet{Weidner2005} and \citet{Pflamm-Altenburg2007} have suggested there may be an upper limit to the maximum stellar mass that can form at low SFRs. In this scenario, the slope of the upper end of the IMF varies as a function of SFR. Using the framework of \citet[][their Figure 11]{Weidner2005}, and \citet[][their Figure 4]{Pflamm-Altenburg2007}, the maximum stellar mass that would be able to form at a SFR $= 10^{-5}$ \msun\ yr$^{-1}$ is $\sim2.5$ \msun. This is an order of magnitude lower than the mass of an $\sim$O8 star ($M_* = 25$ \msun) likely responsible for ionizing the \HII\ region. Indeed, the mere presence of the \HII\ region in Leo~P provides observational evidence that the maximum stellar mass capable of being formed at low SFRs is significantly greater than 2.5 \msun. 

\vspace{-10pt}

\subsection{Variable Star Identification}
We used the multi-epoch photometry to search for short-period variable stars. As
noted in Section~2.2, the spatial cuts were relaxed to include a larger search area 
as RR~Lyrae variable stars can be located in the stellar haloes of galaxies. 
Variable star candidates were identified first using a series of automated cuts, 
similar to those used in the study of IC 1613 \citep{Dolphin2001b}.   Each variable star
was required to meet five criteria testing the photometry, variability, and periodicity.
First, the object had to have well-measured photometery.  While it is possible for a variable 
star to be part of a blend, our PSF-fitting photometry attempts to fit the profile to that 
of a single star and thus is unreliable for measuring blended stars.
The next cuts were intended to eliminate non-variable or weakly-variable stars.  
The simplest requirement was that the RMS scatter of the magnitude measurements,

\begin{align}
& \sigma_{mag} =  (\frac{1}{N_{F475W}+N_{F814W}} \nonumber\\
& * ( \sum_{i=1}^{N_{F475W}} (F475W_i - \overline{F475W})^2 \nonumber\\
& + \sum_{i=1}^{N_{F814W}} (F814W_i - \overline{F814W})^2 ) ) ^{1/2}
\end{align}

had to be 0.14 magnitudes or greater.  Next, the overall reduced $\chi^2$,

\begin{align}
& \chi^2 = \frac{1}{N_{F475W}+N_{F814W}} \nonumber \\
& ( \sum_{i=1}^{N_{F475W}} \frac{(F475W_i - \overline{F475W})^2}{\sigma_i^2} \nonumber \\
& + \sum_{i=1}^{N_{F814W}} \frac{(F814W_i - \overline{F814W})^2}{\sigma_i^2} )
\end{align}

had to be 3.0 or greater.  These two cuts eliminate stars that are weakly variable 
and those for which the photometry signal-to-noise is insufficient to discern variability.  
We also clipped 1/3 of the extreme points and required that the recalculated reduced 
$\chi^2$ exceed 0.5; this was intended to eliminate non-variable stars with a 
few bad points.

The final test was a modified Lafler-Kinman algorithm \citep{Lafler1965}, which 
tests for periodicity.  This was implemented by computing $\Theta$ for periods 
between 0.1 and 4.0 days.  The $\Theta$ parameter is calculated by determining 
the light curve for a trial period and using the equation
\begin{equation}
\Theta = \frac {\sum_{i=1}^{N}(m_i - m_{i+1})^2}{\sum_{i=1}^{N}(m_i - \overline{m})^2},
\end{equation}
where $N$ is the number of exposures for a given filter, $m_i$ is the magnitude 
at phase $i$, and $\overline{m}$ is the mean magnitude.  If the trial period is 
the correct period, each magnitude $m_i$ will be close to the adjacent magnitude 
$m_{i+1}$, giving a value of $\Theta$ that decreases as $1/N^2$.  If the trial 
period is incorrect, there will be less correlation and consequently larger 
values of $\Theta$.  Because we had data in two filters, we combined 
the $\Theta$ values with:
\begin{equation}
\Theta = \frac{4}{ ( 1/\sqrt{\Theta_{F475W}} + 1/\sqrt{\Theta_{F814W}} )^2}.
\end{equation}
For a star to pass the periodicity test, its minimum value of $\Theta$ had to 
be 0.85 or less. For more discussion on the use of the modified Lafler-Kinman algorithm 
to test for periodicity and specifics on the parameters, see \citet{Dolphin2002b, Dolphin2004}. 

\begin{figure}[h]
\includegraphics[width=1.13\columnwidth]{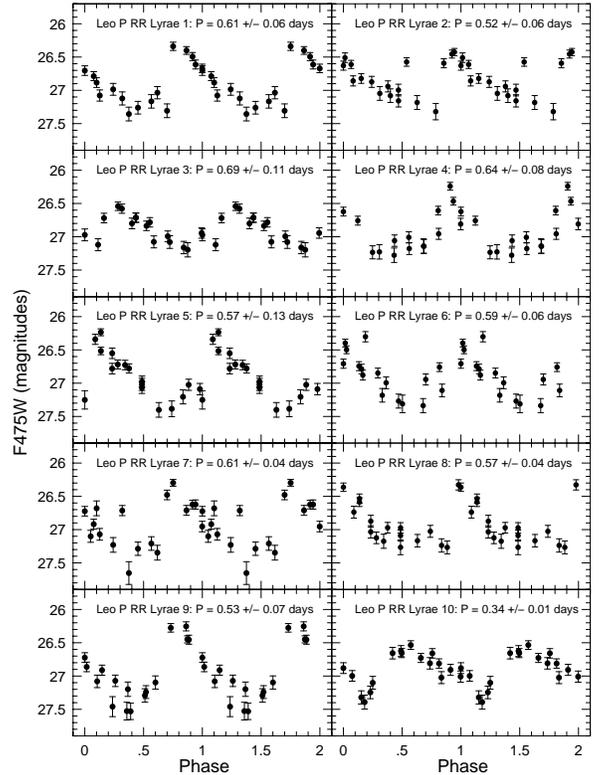}
\vspace{-35pt}
\caption{Light curves for the ten identified RR~Lyrae candidate stars in Leo~P. The measured periods are noted in the individual panels for each star. Complete photometry is available in Table~\ref{tab:RRLyrae_phot}.}
\label{fig:light_curves}
\end{figure}

Our photometry identified 13448 total objects, of which 45 
passed our automatic selection process.  Three of these were clearly
RR Lyrae stars by location in the CMD, light curve shape, and period.
The $\chi^2$ criterion was relaxed, and the search was constrained to an 
area centered on the horizontal branch, and ten more candidates were identified.
Of these new ten, three of these candidate variables appeared to not be 
clean detections upon manual examination of the images and were removed 
from the list resulting in a final total of ten candidate RR Lyrae stars. 
The light curves for the ten variable stars are presented in Figure~\ref{fig:light_curves}; 
complete photometry in the ACS filters is available in Table~\ref{tab:RRLyrae_phot}.

\begin{table}
\begin{center}
\caption{Photometry of RR~Lyrae Candidates}
\vspace{-10pt}
\label{tab:RRLyrae_phot}
\end{center}
\begin{center}
\begin{tabular}{lc}
\multicolumn{2}{c}{\bf Star ID 1} \\
\multicolumn{2}{c}{10:21:44.388 $+$18:05:07.63}\\
\hline
\hline
		&  F475W	\\
Epoch	&  (mag)	\\
\hline 
56770.631305 & 26.70 $\pm$ 0.07 \\
56770.710315 & 27.08 $\pm$ 0.08 \\
56770.829933 & 27.12 $\pm$ 0.10 \\
56770.909546 & 27.26 $\pm$ 0.10 \\
56771.626484 & 27.04 $\pm$ 0.09 \\
\\
\hline
		&  F814W	\\
Epoch	&  (mag)	\\
\hline
56770.647902 & 26.04 $\pm$ 0.10 \\
56770.692595 & 25.95 $\pm$ 0.10 \\
56770.846531 & 26.03 $\pm$ 0.09 \\
56770.891826 & 26.48 $\pm$ 0.15 \\
56771.643081 & 26.25 $\pm$ 0.11 \\
\hline 
\end{tabular}
\end{center}
\tablecomments{Sample photometry of the first five epochs for Star ID 1 in both ACS filters. The RA and Dec (J2000) coordinates for the star are listed at the top. These magnitudes have not been corrected for foreground extinction. The full seventeen epoch photometry for all ten RR~Lyrae candidates is published in its entirety in the electronic edition of this journal. A portion is shown here for guidance regarding its content. Mean magnitudes for each of the ten stars can be found in Table~\ref{tab:RRLyrae_sum}.}
\end{table}

The absolute magnitudes of RR~Lyrae stars have been determined based on the Johnson $BVRI$ filter systems. Therefore, we transform our measured ACS F475W and F814W magnitudes to standard Johnson $V$ and $I$ magnitudes using the calibration from \citet{Saha2011}:

\begin{equation}
V = 0.026 + F475W - 0.406 * (F475W - F814W) \label{eq:V}
\end{equation}
\begin{equation}
I = -0.038 + F814W + 0.014 * (F475W - F814W). \label{eq:I}
\end{equation}

\begin{table*}
\begin{center}
\caption{Summary of RR~Lyrae Candidates in Leo~P}
\label{tab:RRLyrae_sum}
\end{center}
\vspace{-20pt}
\begin{center}
\begin{tabular}{ccccccr}
		& RA			& Dec		& $<$F475W$>$	& $<$F814W$>$	& $<$V$>$	& Period  \\
Star ID	&(J2000)		& (J2000)		& (mag)			& (mag)			& (mag)		& (days)	\\
\hline 
\hline 
01		& 10:21:44.388 & $+$18:05:07.63 & $26.87\pm0.09$ & $26.04\pm0.07$ & $26.56\pm0.14$ & 0.61$\pm0.06$ \\
02		& 10:21:47.080 & $+$18:03:59.16 & $26.85\pm0.08$ & $26.22\pm0.07$ & $26.62\pm0.12$ & 0.52$\pm0.06$  \\
03		& 10:21:44.324 & $+$18:05:33.73 & $26.89\pm0.06$ & $26.11\pm0.07$ & $26.56\pm0.09$ & 0.69$\pm0.11$  \\
04		& 10:21:43.240 & $+$18:05:53.13 & $26.90\pm0.08$ & $26.16\pm0.06$ & $26.62\pm0.12$ & 0.64$\pm0.08$  \\
05		& 10:21:44.722 & $+$18:05:39.91 & $26.90\pm0.10$ & $26.25\pm0.07$ & $26.66\pm0.15$ & 0.57$\pm0.13$  \\
06		& 10:21:42.750 & $+$18:05:46.47 & $26.89\pm0.09$ & $26.17\pm0.05$ & $26.63\pm0.13$ & 0.59$\pm0.06$  \\
07		& 10:21:43.704 & $+$18:05:34.37 & $26.89\pm0.10$ & $26.25\pm0.07$ & $26.66\pm0.15$ & 0.61$\pm0.04$  \\
08		& 10:21:45.227 & $+$18:05:46.86 & $26.91\pm0.07$ & $26.11\pm0.06$ & $26.61\pm0.11$ & 0.57$\pm0.04$  \\
09		& 10:21:44.633 & $+$18:05:21.31 & $26.84\pm0.12$ & $26.27\pm0.06$ & $26.64\pm0.18$ & 0.53$\pm0.07$  \\
10		& 10:21:44.087 & $+$18:06:22.92 & $26.86\pm0.06$ & $26.24\pm0.07$ & $26.63\pm0.10$ & 0.34$\pm0.01$  \\
\\
\multicolumn{3}{c}{\bf Mean Magnitudes} & $26.88\pm0.02$ & $26.18\pm0.07$ & $26.62\pm0.03$ & \\
\hline
\end{tabular}
\end{center}
\tablecomments{10 RR~Lyrae variable star candidates were identified as described in Section~2.4. The coordinates, phase-weighted mean magnitudes, and periods of the individual stars are listed, followed by a mean magnitude computed for all ten stars. These magnitudes have not been corrected for foreground extinction.}
\end{table*}

\noindent These transformations are an improvement over the previous calibrations from \citet{Sirianni2005} as they were derived based on a larger number of stars down to fainter magnitudes and over a greater range in color. Furthermore, the transformations from \citet{Saha2011} were based on updated corrections for time-dependent charge transfer losses in the ACS detector. 

To compute mean magnitudes, we calculated a phase-weighted average using
\begin{equation}
\langle m \rangle = -2.5 \log \sum_{i=1}^{N} \frac{\phi_{i+1} - \phi_{i-1}}{2} 10^{-0.4 m_i},
\end{equation}
where $\phi_i$ is the phase and $m_i$ is the magnitude at each point along the light curve.  
Finally, the mean magnitudes for all ten stars are computed to be F475W $= 26.88\pm0.02$, F814W $=  26.18\pm0.07$, and V $= 26.62\pm0.03$. Note, these mean magnitudes are not corrected for foreground extinction. Table~\ref{tab:RRLyrae_sum} summarizes the mean magnitudes for each star in the three filters and the final mean magnitudes for all ten stars. The location of the RR~Lyrae candidates within Leo~P are shown in Figure~\ref{fig:spatial}.

\begin{figure}[h]
\includegraphics[width=\columnwidth]{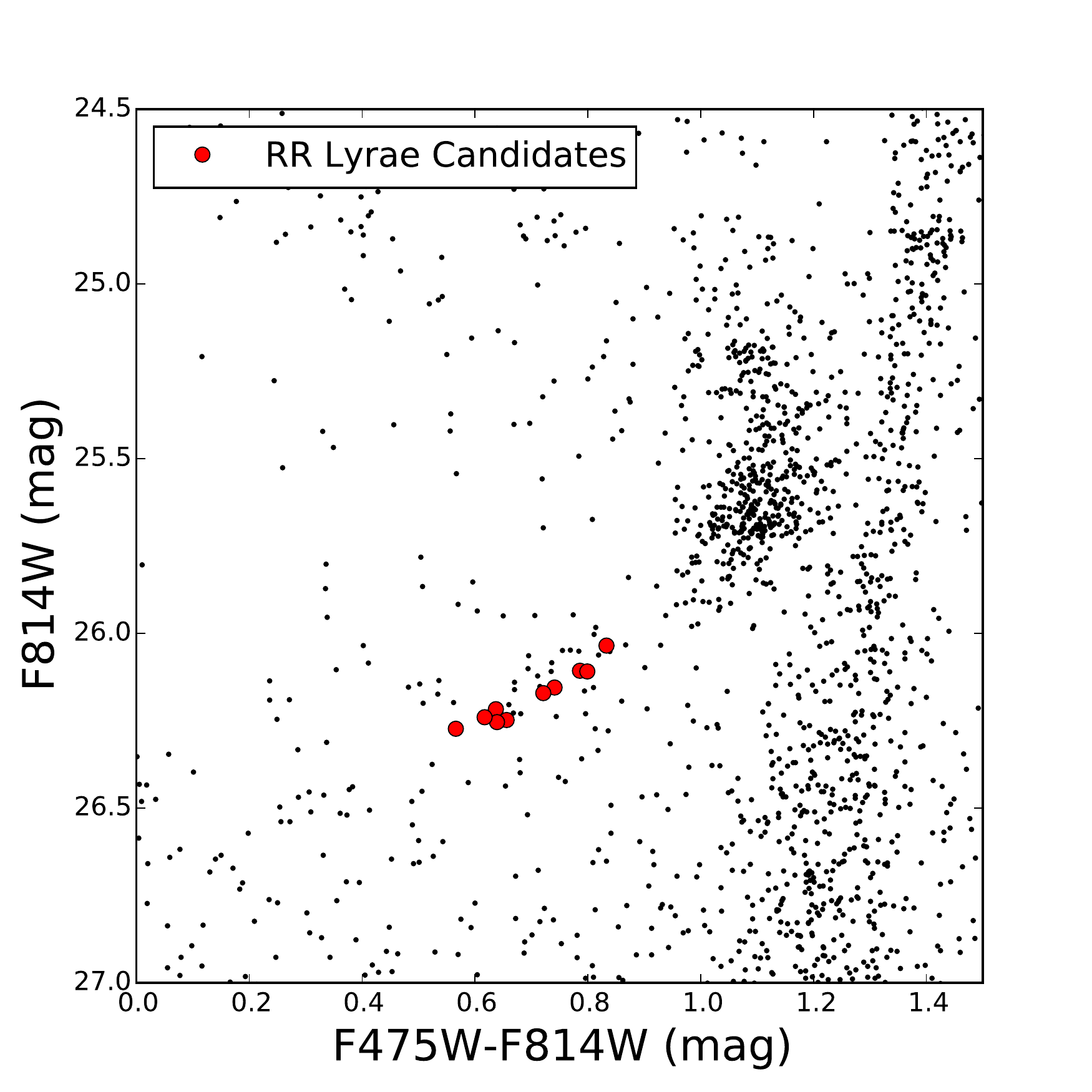}
\caption{A magnified view of the HB and RC region of the CMD. A portion of the RGB is identifiable to the right. The ten RR~Lyrae candidates are plotted in red. As expected, these stars lie in a tight sequence in CMD space.}
\label{fig:cmd_rrlyrae}
\end{figure}

Figure~\ref{fig:cmd_rrlyrae} shows the location of the ten RR~Lyrae candidate variable stars in the CMD. All ten stars were found in a narrow range in luminosity. Because the HB is $\sim$ 2 magnitudes above the 50\% completeness limit for our observations, we likely did not miss any RR Lyrae stars due to depth of photometry.  Also, because of the number and spacing of the epochs, we are likely to have a very high completeness. The data provided excellent coverage for the range of relevant variable star periods. Due to the large number of observational epochs and the fact that the cadence was set to minimize redundancies, the maximum gap in the phase coverage was always less than 0.15 days for periods in the range 0.1 to 1.0 days (see Figure~\ref{fig:light_curves}). Consequently, our mean magnitude measurements are robust, as indicated by the very small dispersion ($\sigma_V = \pm0.03$) in the sample. Thus, the distance to Leo~P derived from the RR Lyrae stars (see Section~3) is limited by the systematic uncertainty of the calibration used.

\section{The Distance to Leo~P\label{distance}}
The previously published distance modulus measurement of $26.19^{+0.17}_{-0.50}$ mag from \citet{McQuinn2013} was determined by applying the TRGB distance method to the photometry of the resolved stellar population from LBT V and I band imaging. Despite photometry reaching 3 magnitudes below the TRGB, the distance uncertainties in the ground-based observations are large due to the small number of stars in the RGB. Using Monte Carlo simulations of synthetic stellar populations analogous to low-mass galaxies, \citet{McQuinn2013} showed that the measured break in the LBT I band LF at the top of the RGB may, in fact, be below the actual TRGB luminosity. The large lower uncertainty estimated from the simulations put constraints on the actual TRGB of up to 0.5 mag brighter than the measured value. Using the same TRGB distance method described in \citet{McQuinn2013}, we re-measured the TRGB distance to Leo~P using the HST data. This exercise yielded an identical measurement of the distance modulus of 26.19 mag as the LBT data, but does not mitigate the uncertainties as measured by the Monte Carlo simulations.

However, the new HST data reach 2 mag below the HB, enabling two additional distance techniques to be applied. First, as described above, the cadence of our observations was designed to maximize identification of RR~Lyrae stars, thus enabling a distance measurement based on the average V-band luminosity of this class of stars. Second, the luminosity of the HB provides a means to measure the distance based on a standard candle approach. The HB luminosity can be more robustly measured than the TRGB with sparse data as the mean luminosity of the HB is not as impacted by the presence or absence of a few individual stars. 

\subsection{RR~Lyrae Distance Measurement}
As discussed above, the mean V magnitude of the ten RR~Lyrae stars is $26.62\pm0.03$ mag. To account for foreground extinction, we subtract $A_V = 0.071$ mag based on the dust maps of \citet{Schlegel1998} with the re-calibration from \citet{Schlafly2011}.

There has been significant work done to calibrate the luminosity of the RR~Lyrae stars as a function of metallicity, however the zero-point can still vary by $\sim0.1$ mag between calibrations \citep[for detailed reviews, see][among others]{Smith1995, Cacciari2003, Sandage2006, Catelan2009}. Some of the differences are due to the difficulty quantifying the luminosity dependence on stellar metallicity. Regardless, the choice of calibration introduces a systematic error to our distance measurement, dominating the uncertainties. We use the calibration from \citet{Carretta2000} which was derived using dozens of RR~Lyrae stars in globular clusters and anchored to several independent distance measurements:

\begin{equation}
M_V = (0.18\pm0.09) * (\rm{[Fe/H]} + 1.5) + (0.57\pm0.07). \label{eq:calib1_RRLyrae}
\end{equation}

\noindent As seen in Equation~\ref{eq:calib1_RRLyrae}, the RR~Lyrae luminosity depends on the metallicity of the population. For an estimate of the metallicity of the oldest stars, we fit Padova-Trieste stellar evolution isochrones \citep[PARSEC;][]{Bressan2012, Chen2014, Tang2014} to the RGB. The PARSEC models are the result of a thorough revision of the previous Padova stellar evolution code with the most up to date input physics. The PARSEC library also includes post-helium core flash phases of evolution and has been adjusted to a revised solar abundance. In Figure~\ref{fig:cmd_highlights}, we overlay isochrones for a 12 Gyr old population with [Fe/H] values of $-1.9, -1.8$, and $-1.7$. The best-fitting isochrone to the mean color of the RGB has a metallicity value of [Fe/H] $= -1.8\pm0.1$; the uncertainty is based on the metallicities of the same age isochrones that span the approximate width of the RGB. The $-1.8$ value seems reasonable as it is lower than the present day gas-phase oxygen abundance measurement of $-1.52$ (assuming [O/Fe] $\sim0$), and consistent with stellar metallicities spectroscopically measured in low-mass dwarfs \citep[e.g.,][]{Kirby2010}. We experimented with using the Dartmouth stellar evolution isochrones which are known to provide a good fit to the RGB in low-metallicity systems. The best-fitting Dartmouth isochrones had a similar metallicity value and range of $-1.9\pm0.1$, providing a consistency check on the values from the PARSEC isochrones. Using an [Fe/H] value of $-1.8\pm0.1$ in Equation~\ref{eq:calib1_RRLyrae} yields $M_V = 0.52\pm0.20$, corresponding to a distance modulus of $26.04\pm0.21$ mag.

As noted above, there has been significant work done to calibrate the luminosity of the RR~Lyrae stars. For comparison, we consider two additional calibrations from the literature. First, we utilize the calibration from \citet{Clementini2003} based on photometry and spectroscopy of more than 100 RR~Lyrae stars in the Large Magellenic Cloud:

\begin{equation}
M_V = (0.866\pm0.085) + (0.214\pm0.047) * \rm{[Fe/H]}. \label{eq:calib2_RRLyrae}
\end{equation}

\noindent Using the same [Fe/H] value of $-1.8$, Equation~\ref{eq:calib2_RRLyrae} yields $M_V = 0.48\pm0.12$, corresponding to a distance modulus of $26.07\pm0.13$ mag.

Second, we consider the calibration from \citet{Bono2003} with uses an updated theoretical prescription of the pulsating modes of RR~Lyrae stars:

\begin{equation}
M_V = (0.718\pm0.072) + (0.177\pm0.069) * \rm{[Fe/H]}. \label{eq:calib3_RRLyrae}
\end{equation}

\noindent This calibration yields an absolute magnitude of $M_V = 0.40\pm0.15$, corresponding to a distance modulus of $26.16\pm0.15$ mag. The distances from these additional calibrations agree within the uncertainties to the distance calculated based on the calibration from \citet{Carretta2000} above in Equation~\ref{eq:calib1_RRLyrae}.

\begin{table*}
\begin{center}
\caption{Distance Measurements to Leo~P}
\label{tab:distances}
\end{center}
\vspace{-20pt}
\begin{center}
\begin{tabular}{lrrr}
Method			& (m-M)$^o$				& Distance (Mpc)		& Calibration \\
\hline 
\hline 
TRGB (LBT)		& 26.19$^{+0.17}_{-0.50}$	& 1.72$^{+0.14}_{-0.40}$	& \citet{Rizzi2007, Carretta2000}\\
TRGB (HST)		& 26.19$^{+0.17}_{-0.50}$	& 1.72$^{+0.14}_{-0.40}$	& \citet{Rizzi2007, Carretta2000}\\
{\bf Horizontal Branch}& {\bf 26.05}${\bf \pm0.20}$	& {\bf 1.62}${\bf \pm0.15}$&  \citet{Carretta2000}\\
{\bf RR~Lyrae} 		& {\bf 26.04}${\bf \pm0.21}$	& {\bf 1.61}${\bf \pm0.16}$& \citet{Carretta2000}\\
RR~Lyrae 		& 26.07$\pm0.13$			& 1.64$\pm0.10$		& \citet{Clementini2003}\\
RR~Lyrae 		& 26.16$\pm0.15$			& 1.70$\pm0.12$		& \citet{Bono2003}\\
\hline
\end{tabular}
\end{center}
\tablecomments{TRGB distance was measured from LBT optical imaging of the resolved stellar populations \citep{McQuinn2013} and confirmed via the HST data presented here. For the TRGB distance measured from the HST data, we adopt the  uncertainties from the Monte Carlo simulations in \citet{McQuinn2013}. All values include a correction for foreground extinction of $A_V = 0.071$ mag. For the final distance measurement to Leo~P, we adopt the HB distance based on the calibration by \citet{Carretta2000} because this measurement agrees with the RR~Lyrae distance based on the same calibration but has a slightly lower uncertainty. See Section~3.3 for discussion.}
\end{table*}

\vspace{10pt}

\subsection{HB Distance Measurement}
The luminosity of the red HB stars can also be used as a distance indicator. Similar to the RR~Lyrae stars, the calibration for the HB stars has been done in the Johnson filters. Thus, we use the same transformations described in Equations~\ref{eq:V} and \ref{eq:I} to convert the ACS magnitudes to V and I band magnitudes. To measure the luminosity of the HB, we fit a parametric LF to the observed distribution of stars in the magnitude and color range of the HB feature, without any correction for foreground extinction. This is a similar maximum likelihood approach used to measure the TRGB in many systems \citep[e.g.,][]{Sandage1979, Mendez2002, Makarov2006, Rizzi2007}, based on a probability estimation that takes into account photometric error distribution and completeness using artificial star tests. We assumed the following form for the theoretical LF:

\begin{equation}
 P = exp^{( A*(V-V_{HB}) + B )} + exp^{( -0.5 * ((V-V_{HB})/C)^2 )}
\end{equation}

\noindent where A, B, and C are free parameters. We selected stars in the HB identified by eye in the CMD to use in the fit. Specifically, stars  in the V mag range of $25 - 28$ and within a V$-$I color range of $0.2 - 0.6$ were used. The HB luminosity was measured to be V$ = 26.62\pm0.01$ mag, from which we subtracted $A_V = 0.071$ mag to correct for extinction. 

We use the HB calibration from \citet{Carretta2000} to determine the absolute magnitude of the HB:

\begin{equation}
M_V = (0.13\pm0.09)  * (\rm{[Fe/H]} + 1.5) + (0.54\pm0.04) \label{eq:calib_HB} \\
\end{equation}

\noindent In the same manner described for the RR~Lyrae stars, we assume an [Fe/H] value of $-1.8\pm0.1$. The HB luminosity based on the above calibration yields $M_V = 0.50\pm0.20$, in excellent agreement with the range in calibrated magnitudes of the ten RR~Lyrae stars. Based on this HB luminosity, the distance modulus is $26.05\pm0.20$.

\subsection{Adopted Distance Measurement}
Table~\ref{tab:distances} summarizes the distance measurements to Leo~P based on the TRGB from the LBT and HST data sets and RR Lyrae and HB measurements from the HST data. The distance moduli determined from the different calibrations and techniques agree within the uncertainties. The calibrations from \citet{Carretta2000} provide consistency across the RR~Lyrae, HB, and TRGB distance techniques, allowing us to compare the three methods without introducing a systematic uncertainty. These three techniques yield distance moduli of $26.04\pm0.21$, $26.05\pm0.20$, and $26.19^{+0.17}_{-0.50}$ respectively. The consistency of the RR~Lyrae and HB distance measurements reinforces the conclusion that the TRGB identified in the CMD from \citet{McQuinn2013}, similarly to the $HST$ data presented here, is indeed a false tip. Furthermore, all of the distance measurements are within the uncertainties estimated via Monte Carlo simulations of synthetic populations from \citet{McQuinn2013} supporting this approach for determining uncertainties on the TRGB in this very low-mass regime. 

The HB and RR~Lyrae distances based on the calibration from \citet{Carretta2000} are very similar. We adopt the distance modulus of $26.05\pm0.20$ mag from the HB stars due to the slightly lower uncertainty, yielding a distance of $1.62\pm0.15$ Mpc. The refined distance places Leo~P in the NGC~3109 association of dwarf galaxies \citep{Tully2006} including Antlia, Sextans~A, and Sextans~B as proposed by \citet{McQuinn2013}. The closest of these galaxies is Sextans~B at a distance of $\sim0.4$ Mpc. In Table~\ref{tab:properties}, we include updated distance dependent parameters using this revised distance measurement.

\section{Star Formation History Methodology\label{sec:sfh}}
The SFH was measured using the numerical CMD fitting program {\tt MATCH} \citep{Dolphin2002a}. Briefly, {\tt MATCH} uses a prescribed IMF and stellar evolutionary isochrones to create a series of synthetic simple stellar populations (SSPs) of different ages and metallicities. The synthetic SSPs are modeled using the photometry and recovered fractions of the artificial stars as primary inputs. The modeled CMD that best-fits the observed CMD based on a Poisson likelihood statistic provides the most likely SFH of the galaxy. 

The solutions are based on a Kroupa IMF \citep{Kroupa2001} and the newly available PARSEC models \citep{Bressan2012, Chen2014, Tang2014}. We assume a binary fraction of 35\% with a flat secondary distribution. While SFH solutions have been shown to be fairly insensitive to the choice of binary fraction, assumed fraction values between one and two thirds tend to improve the overall fit \citep{Monelli2010a}. Extinction is a free parameter fit by {\tt MATCH}. Both foreground and internal extinction can broaden the features in a CMD, but they are expected to be low for Leo~P based on its high Galactic latitude and low metallicity. Foreground extinction is estimated to be $A_g=0.086$ mag \citep{Schlafly2011} compared with the best-fitting extinction value of $A_{F475W} = 0.12\pm0.05$, corresponding to both foreground and internal extinction which is taken into account in the SFH solution. The general agreement between the foreground extinction from \citet{Schlafly2011} and the best-fitting extinction value from {\tt MATCH} provides an additional consistency check on the SFH solution. Because the photometric depth of the data does not fully constrain the metallicity evolution of the systems, we assumed that the chemical enrichment history, $Z(t)$, is a continuous, non-decreasing function over the lifetime of the galaxy. Uncertainties on the SFHs take into account both systematic uncertainties from the stellar evolution models \citep{Dolphin2012} and random uncertainties due to the finite number of stars in a CMD \citep{Dolphin2013}. 

\begin{figure}[ht]
\includegraphics[width=0.48\textwidth]{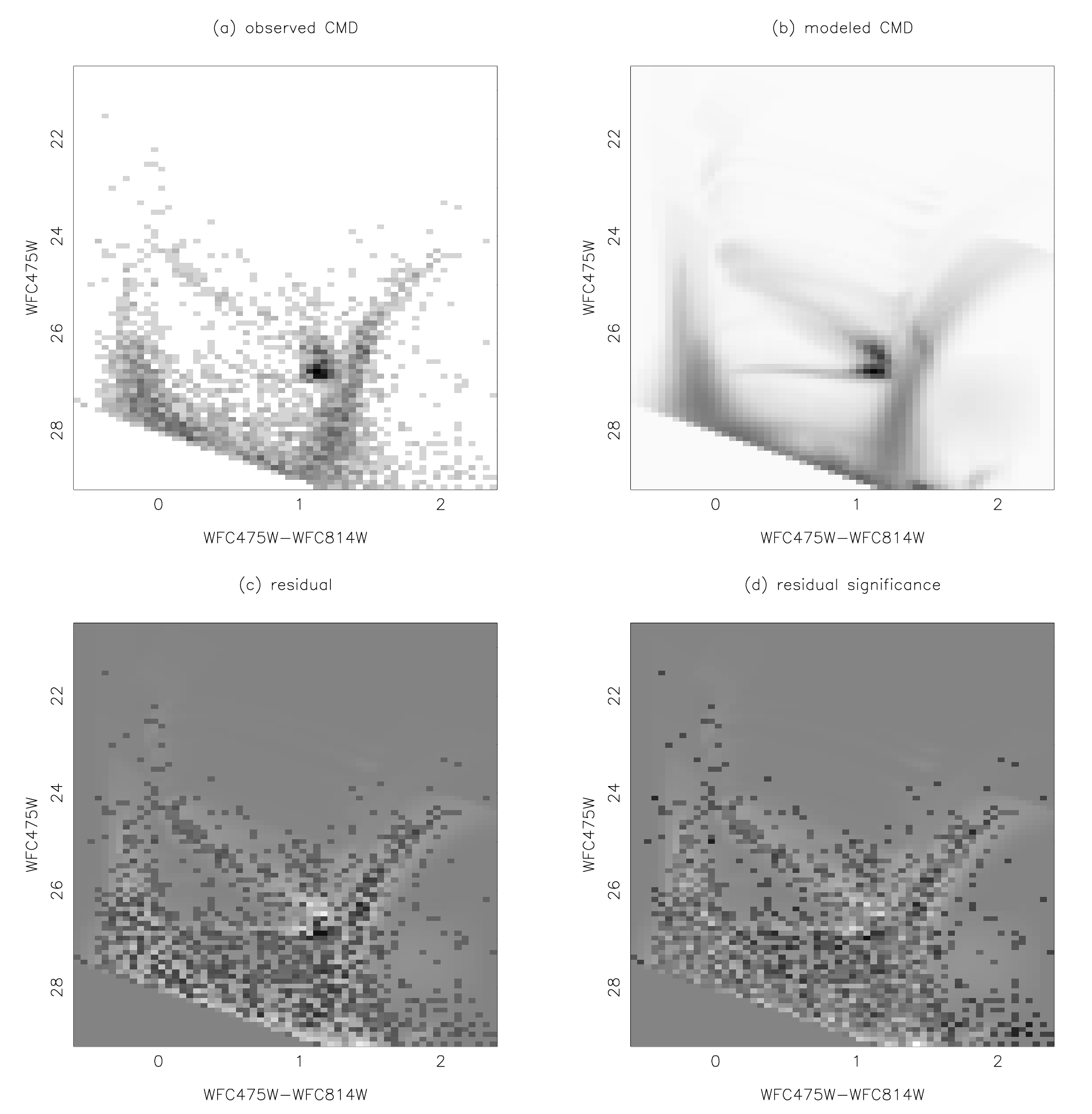}
\caption{The observed and modeled CMD for Leo~P. (a) Observed Hess diagram of Leo~P from photometry of the $HST$ images. (b) Modeled Hess diagram based on the best-fitting SFH solution to the observations with the PARSEC library. For both (a) and (b) the grayscale is based on the number of stars in each bin. (c) Residual Hess diagram between observed and modeled CMD of Leo~P. (d) Residual significance between the observed and modeled CMD of Leo~P. The grayscale spans 5$\sigma$ and is based on the significance of each bin in the residual (panel c) relative to the standard deviation of a Poisson distribution. The final panel (d) highlights the good agreement between the modeled CMD and observed CMD.}
\label{fig:model_CMD}
\end{figure}

Distance is a free parameter fit by {\tt MATCH}. {\tt MATCH} can be run without distance constraints to determine the best-fitting distance to the CMD by the models. This distance can be compared with the measured distance from an independent approach as a consistency check on the solutions. The final SFH solution is derived by fixing the distance to be the best measured value. In the case of Leo~P, the best-fitting distance modulus by the PARSEC models is 26.15$\pm0.05$ which is in good agreement with the measured values in Table~\ref{tab:distances}. We experimented by fixing the distance modulus to the different values derived from the HB and RR~Lyrae stars. These tests showed that the modeled CMD with a smaller adopted distance modulus of 26.05 is an improvement over a larger distance modulus of 26.19. The final SFH solution for each model is based on fixing the distance modulus to the adopted value of 26.05. 

\begin{figure*}[ht]
\includegraphics[width=\textwidth]{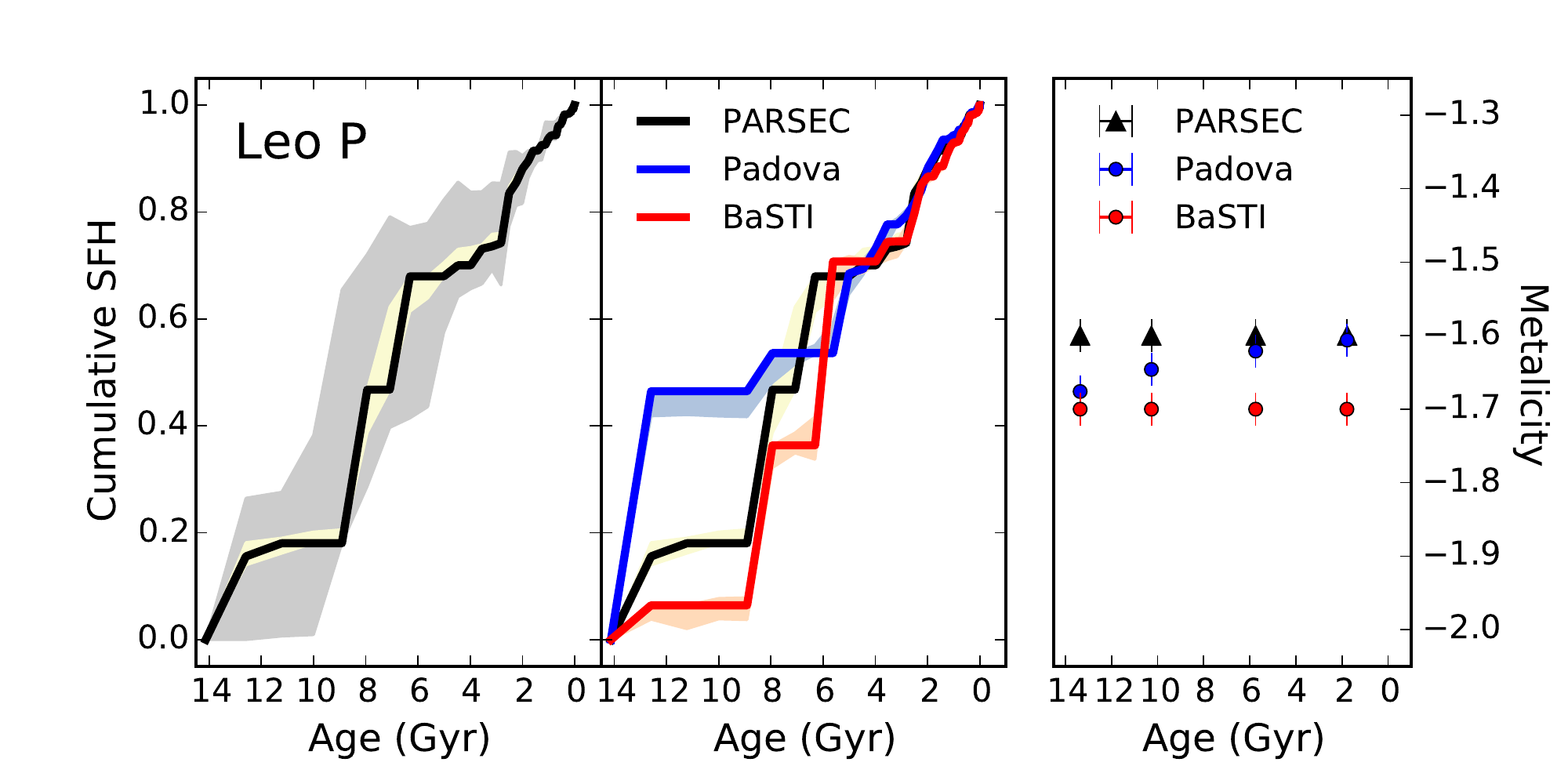}
\caption{\textit{Left panel:} Best-fitting cumulative SFHs for Leo~P derived using the PARSEC stellar evolution libraries. Random uncertainties are plotted in yellow \citep{Dolphin2013}; combined random and systematic uncertainties are plotted in grey \citep{Dolphin2012}. \textit{Middle panel:} For comparison with the PARSEC models, we present the best-fitting cumulative SFHs using two additional stellar libraries:  Padova libraries (blue), and BaSTI libraries (red). Random uncertainties are plotted in shaded colors for each library. The systematic uncertainties derived for the PARSEC libraries using \citep{Dolphin2012} (shown in the left panel) are supported by this comparison. \textit{Right panel:} Chemical evolution [M/H] solution for Leo~P including random uncertainties. For clarity, we plot limited modeled values of stellar metallicity during periods of star formation. The chemical enrichment solutions show good agreement  ranging from [M/H] of $-1.6$ to $-1.7$ across all ages.}
\label{fig:sfh_metals}
\end{figure*}

In Figure~\ref{fig:model_CMD}, we present the observed CMD, modeled CMD using the PARSEC isochrones, residual CMD, and residual significance from {\tt MATCH}. The modeled CMD is an excellent fit to the observed CMD of Leo~P, with a corresponding $\chi^2$ value of 1.26. The residuals are most notable in some of the typical areas including the width of the RGB and the RC. While small, the residuals highlight the differences between the observations and the model from the isochrones. The differences in the SFH implied by the residuals are well within the measured uncertainties and do not impact our overall conclusions. 

The SFH solution depends on the ability of the stellar evolution isochrones to accurately model the luminosity, color, and number density of the features in a CMD \citep[see][for a comprehensive review]{Gallart2005}. Thus, as an additional test, we used two different stellar evolution models to derive the SFHs: the Padua stellar evolution models \citep{Marigo2008} with updated AGB tracks from \citet{Girardi2010} and the BaSTI models \citep{Pietrinferni2004}. The resulting CMDs modeled from the Padova and BaSTI libraries were also a good match to the observed CMD, with slightly larger discrepancies in the RGB and RC regions. Specifically, the Padova models expected RGB stars that were bluer than observed, while both libraries produce more HB stars than observed. The corresponding $\chi^2$ values for the Padova and BaSTI libraries were 1.34 and 1.37 respectively. Note that the BaSTI models do not include stars with ages $\ltsimeq30$ Myr. We present the SFH results for all three libraries in the following section.

\section{The Star Formation History of Leo~P}

The left panel of Figure~\ref{fig:sfh_metals} presents the cumulative SFH of Leo~P based on the best-fitting modeled CMD from {\tt MATCH} using the PARSEC models with both random and systematic uncertainties. The fraction of stellar mass formed at each epoch is listed in Table~\ref{tab:sfh}. Comparison solutions derived using the Padova and BaSTI libraries are shown the middle panel, with constraints on the chemical evolution shown in the right panel. Given the range in the three solutions, the SFH is consistent with a relatively constant level of star-formation activity over cosmic timescales as seen in more massive dIrrs \citep[e.g., IC~1613;][]{Skillman2014}. The best-fitting solutions for Leo~P also suggest that star formation may have been damped after early epochs (i.e., post-reionization), with little SF occurring between $8-12$ Gyr ago. As seen in the middle panel of Figure~\ref{fig:sfh_metals}, the solutions from the three stellar libraries provide a range for the degree and duration of damping. Although different in amplitude, this is reminiscent of the delayed star formation seen in the SFH of the gas-rich LG dwarfs Leo~A \citep{Cole2007} and DDO~210 \citep{Cole2014}. Regardless, from the SFH it is clear that Leo~P experienced star formation at $t\geq10$ Gyr, in agreement with the identification of ten RR~Lyrae stars. Over the last $\sim4$ Gyr, the solutions from all three libraries show continual SF at an approximately constant rate. 

The stellar mass of Leo~P can be estimated by integrating the SFH over time and assuming a 30\% recycling fraction \citep{Kennicutt1994}. We use the PARSEC solution for this calculation and find the present day stellar mass in Leo~P to be $5.6^{+0.4}_{-1.9}\times10^5$ \msun, reported in Table~\ref{tab:properties}. The resulting stellar mass-to-light ratio is $M_*/L_V = 1.25$, based on a solar $M_V = 4.86\pm0.02$ \citep{Pecaut2013}. The stellar mass derived from the SFH is in good agreement with the previous estimate of $5.7^{+0.4}_{-1.8}\times10^5$ \msun\ from ground-based LBT images \citep{McQuinn2013} which used the M/L ratio formalism from \citet{Bell2001}.

\begin{table}
\begin{center}
\caption{Cumulative SFH of Leo~P}
\label{tab:sfh}
\end{center}
\vspace{-15pt}
\begin{center}
\begin{tabular}{cc}
\multicolumn{2}{c}{Fraction of Stellar Mass}\\
 \multicolumn{2}{c}{Formed by each Epoch} \\
\hline 
\hline 
f$_{10.10 }$ &  0.16$^{+0.11}_{-0.16}$ \\
f$_{10.05 }$ &  0.18$^{+0.10}_{-0.17}$ \\
f$_{10.00 }$ &  0.18$^{+0.20}_{-0.17}$ \\
f$_{9.95 }$ &  0.18$^{+0.47}_{-0.00}$ \\
f$_{9.90 }$ &  0.47$^{+0.25}_{-0.18}$ \\
...		& ...	\\
f$_{6.80 }$ &  1.00$^{+0.00}_{-0.00}$ \\
f$_{6.75 }$ &  1.00$^{+0.00}_{-0.00}$ \\
f$_{6.70 }$ &  1.00$^{+0.00}_{-0.00}$ \\
f$_{6.65 }$ &  1.00$^{+0.00}_{-0.00}$ \\
f$_{6.60 }$ &  1.00$^{+0.00}_{-0.00}$ \\
\hline
\end{tabular}
\end{center}
\tablecomments{The fraction of stellar mass formed prior to each log time bin based on the best-fitting SFH using the PARSEC stellar evolution library. Uncertainties include both random and systematic uncertainties. The SFHs are derived assuming the fraction of stellar mass formed is zero at log(t) $= 10.15$ and unity at log(t) $= 6.6$. Integrating the SFH over all time bins, the total stellar mass {\it formed} is $8.6\times10^5$ \msun. Assuming a 30\% recycling fraction, the present day stellar mass in Leo~P is $5.6^{+0.4}_{-1.9}\times10^5$ \msun. The full cumulative SFH is published in its entirety in the electronic edition of this journal. A portion is shown here for guidance regarding its form and content.}
\end{table}

Regardless of the model, the best-fitting chemical evolution histories show very little enrichment in the stars with [M/H] values ranging from $-1.7$ to $-1.6$. These values are close to our estimate of [Fe/H] of $-1.8\pm0.1$ from PARSEC isochrone fits for the oldest stars used in the distance calibrations of the HB and RR~Lyrae stars. Relatively slow chemical evolution has been noted in other low-mass galaxies such as Leo~A, IC~1613, and DDO~210 \citep{Cole2007, Skillman2014, Cole2014}, but Leo~P shows less chemical evolution than any of these systems. A follow-up investigation of the chemical evolution history of Leo~P will be presented in a future paper (K.~B.~W.~McQuinn et al. in preparation).

\section{Leo~P as a Probe of Evolutionary Scenarios}

\subsection{Comparison Sample and Present-Day Luminosities}

To provide context for Leo~P's properties, we compile a comparison sample of very low-mass, low-luminosity galaxies in the nearby universe, listed in Table~\ref{tab:low_mass}. By both necessity and design, the sample is heterogeneous, including both gas-poor dSphs satellites of the Milky Way and gas-rich dIrrs that lie outside the virial radius of the Milky Way but still inside the LG zero velocity boundary. On the one hand, there are very few known {\it gas-rich, star-forming} galaxies in this mass range, despite predictions that they should be the most common galaxy structure in the nearby universe \citep[e.g.,][]{Haynes2011}. As noted in the Introduction, these galaxies have proven elusive observationally.  Thus, to build a larger sample of galaxies in this mass regime, we must include some of the {\it gas-poor} low-mass dSphs detected in close proximity to the Milky Way. On the other hand, the dSphs present very different environment-driven histories than Leo~P and other dIrrs. Thus, these galaxies provide a contrast to evolution in isolation allowing us to explore possible environmental differences on evolution in this mass regime.  

As an extension to our comparison, we also include three dSphs with even lower-lumnosities. The discovery and study of a population of very low-mass Milky Way satellite dSph galaxies, colloquially called ``ultra-faint dwarf'' galaxies, have extended the constraints of structure formation and evolution in the universe. These galaxies are very low-luminosity \citep[$M_V \gtsimeq -8$;][]{Martin2008} with little gas content and a predominantly old, metal-poor, stellar population \citep[e.g.,][]{Sand2009, Sand2010,  Okamoto2008, Okamoto2012, deJong2008b, Hughes2008, Martin2008, Frebel2010, Norris2010, Kirby2010, Kirby2011a, Brown2014, Weisz2014a}. The mass-to-light ratios of these systems indicate that they are dark matter dominated \citep[i.e., M/L $\gtsimeq100$][]{Kleyna2005, Munoz2006, Martin2007, Simon2007}, setting them apart from globular clusters of similar luminosity. Discovered from slight over-densities in the SDSS data set, it appears that these very low surface brightness galaxies are an extension of dSphs to lower masses \citep{Belokurov2007, Clementini2012}.

\begin{table}
\begin{center}
\caption{Comparison of Leo~P with Nearby Very Low-Mass Galaxies}
\label{tab:low_mass}
\end{center}
\vspace{-25pt}
\begin{center}
\begin{tabular}{lrrrr}
				& $M_V$ 		& Dist.	& $M_*$ 			&		\\
Galaxy			& (mag)		& (kpc)	& ($10^5$ \msun)	& Ref.	\\
\hline 
\hline 
\multicolumn{5}{c}{\bf Gas-Rich Low-Mass Dwarfs}\\
Leo~A                   	& $-12.1$        	& 798	& 60 				& 1   		\\
DDO~210			& $-10.6$		& 977	& 16 				& 2		\\
Leo~P			& $-9.27$		& 1640	& 5.6 			& \nodata	\\
Leo~T                   	& $-8.0$          	& 417	& 1.4 			& 1,3        \\
\multicolumn{5}{c}{\bf Low-Mass dSphs}\\
Draco                   	& $-8.8$          	& 76		& 2.9 			& 1,4       	\\
Ursa~Minor              	& $-8.8$          	& 76		& 2.9 			& 1,5        	\\
Canes Venatici I	& $-8.6$		& 218	& 2.3 			& 1,4  	\\
\multicolumn{5}{c}{\bf Very Low-Mass dSphs}\\
Hercules			& $-6.6$		& 132	& 0.37 			& 1,4 	\\
Leo~IV			& $-5.0$		& 154	& 0.19 			& 1,6	 	\\
Canes Venatici II	& $-4.9$		& 160	& 0.079 			& 1,4 	\\
\hline
\end{tabular}
\end{center}
\tablecomments{All values for Leo~P are from this work. Stellar Masses for the comparison sample are based on a stellar mass-to-light ratio of unity.}
\vspace{10pt}
\textbf{References:} (1) \citet{McConnachie2012}; (2) \citet{Cole2014}; (3)  \citet{deJong2008a}; (4) \citet{Martin2008};  (5) \citet{Irwin1995}; (6) \citet{deJong2010}.
\end{table}

From Table~\ref{tab:low_mass}, the present-day luminosities of the dIrrs and dSphs span close to four magnitudes; this range increases to seven magnitudes when including the very low-mass dSphs. The dSphs are all located within $\sim200$ kpc of the Milky Way, well inside the virial radius. The three comparison dIrrs reside outside of this boundary, but still inside the LG \citep{McConnachie2012}.

\begin{figure*}[ht]
\includegraphics[width=\textwidth]{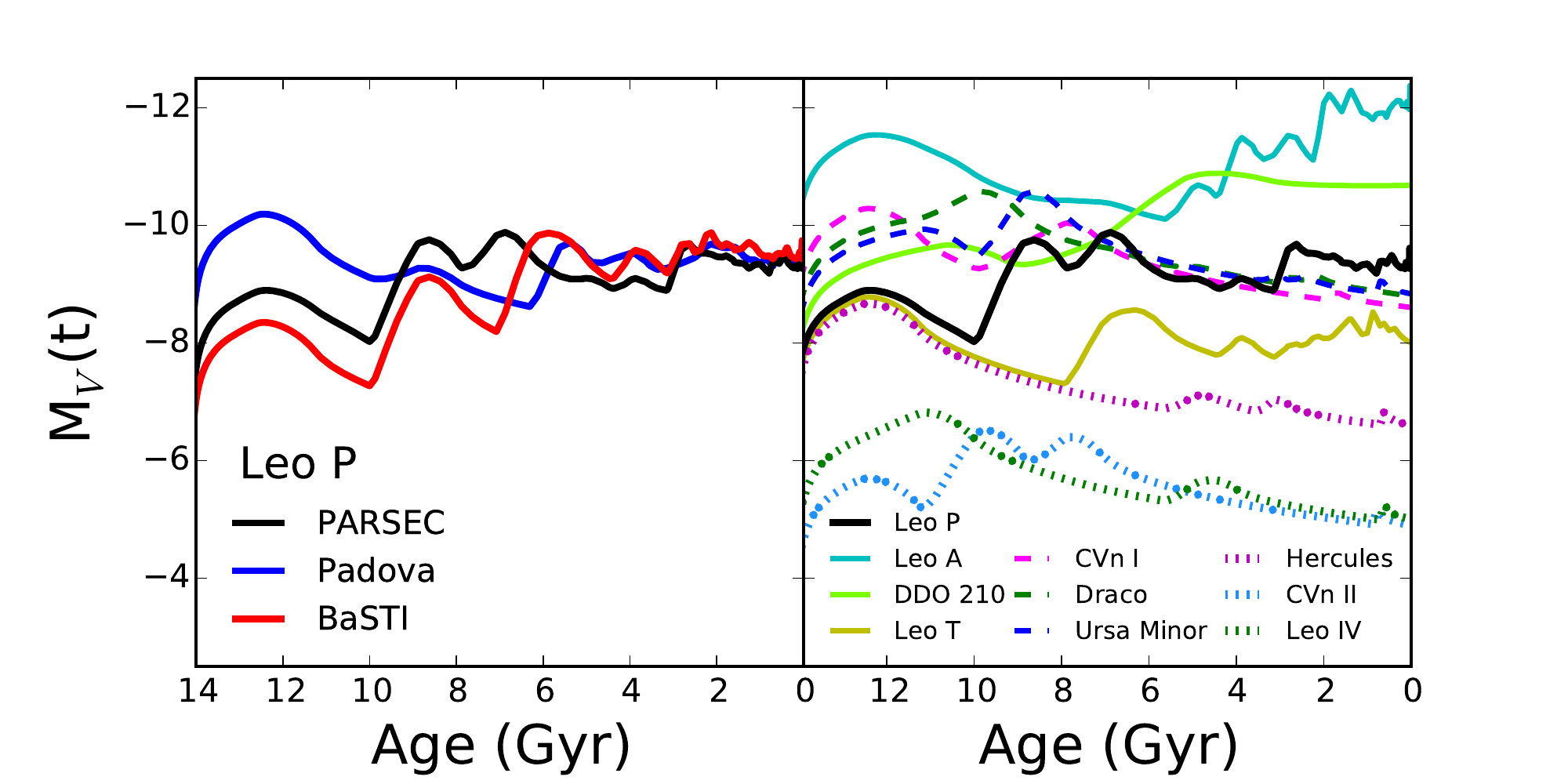}
\caption{The luminosity evolution of Leo~P and nine low-luminosity Milky Way satellites based on their SFHs and stellar population synthesis modeling. The luminosity of Leo~P at early times is consistent with a number of the now gas-poor dSphs which have since faded to lower luminosities due to their lack of star formation. Leo~P is analogous to a dSph evolving in isolation, highlighting the dramatic effect environment has on very low-mass galaxies. Note the smoother profile of DDO~210 is due to the larger time binning of the SFH from \citet{Cole2014}.}
\label{fig:lum_time}
\end{figure*}

\subsection{Luminosity Evolution of Very Low-Mass Galaxies}
Another way to compare these low-mass galaxies is to consider the evolution of their luminosities with time based on their SFHs \citep[e.g.,][]{Weisz2014b}. This can allow one to compare the initial properties of the sample and study whether subsequent changes are correlated with environment. Such a comparison can be made by converting the SFHs into luminosity evolution profiles. The SFHs from resolved stellar populations of various low-mass galaxies have been published by a number of authors \citep[e.g.,][among others]{Cole2007, Kirby2011a, Clementini2012, Brown2014, Weisz2014a, Cole2014}. 

Because our analysis requires tabulated results of the fraction of stars as a function of time, we utilize the results on eight galaxies in our comparison sample derived from $HST$ optical imaging from \citet{Weisz2014a}, and similarly obtained results on one galaxy (DDO~210) from \citet{Cole2014}. The SFH solutions from \citet{Weisz2014a} were derived using the Padova isochrones, and the solution for DDO~210 from \citet{Cole2014} was derived using the PARSEC isochrones. However, the CMDs in both of these previous studies have photometric depths reaching below the old main sequence turn-off (oMSTO), which provides robust constraints on the ancient star formation. SFH solutions derived from different stellar evolution libraries using data with photometry extending below the oMSTO have shown excellent agreement with each other \citep[e.g.,][]{Gallart2005, Monelli2010a, Monelli2010b, Hidalgo2011, Weisz2011}. Thus, the systematic uncertainties between members of our comparison sample should be small. For completeness in our qualitative comparison, we also include the luminosity evolution of Leo~P based on the SFH from the Padova, PARSEC, and BaSTI models, which provides a sense of the systematic uncertainties.

The SFHs were used as input to the stellar population synthesis technique from \citet{Bruzual2003} which builds a synthetic galaxy spectrum as a function of time, assuming a constant metallicity and an IMF.  Because the SFHs from \citet{Weisz2014a} and \citet{Cole2014} were derived from $HST$ data which do not encompass the full optical disks of the galaxies, a correction is necessary. While the SFHs are generally assumed to be representative of the galaxy as a whole, the stellar mass only traces the stars (and luminosity) in the field of view.  As a result, the modeled present-day luminosities from the SFHs are less than the measured luminosities. The exception to this is the luminosity for Leo~P whose stellar disk is contained within the observational footprint. To account for the disparate fields of view between SDSS imaging used to determine the luminosities and data used to measure the SFHs from \citet{Weisz2014a} and \citet{Cole2007}, we normalized the results from \citet{Bruzual2003} by the present-day measured luminosity. This scaled the modeled luminosities while preserving relative changes as a function of time. 

In the left panel of Figure~\ref{fig:lum_time}, we present a comparison of the luminosity evolution ($M_V(t)$) for Leo~P based on the three stellar evolution libraries. Because the best-fitting solution from the Padova models has 50\% of the stellar mass formed in the earliest epochs, compared to 10\% $-$ 20\% from the BaSTI and PARSEC libraries respectively, the Padova luminosity profile also shows the highest luminosity at earliest times. Regardless of the absolute magnitude, all three solutions show a peak in brightness at early times followed by a slow fading in luminosity. 

The right panel of Figure~\ref{fig:lum_time} presents a comparison of $M_V(t)$ for Leo~P using the PARSEC models with six gas-poor dSphs and three gas-rich dIrrs. We chose to highlight the Leo~P solution using the PARSEC models which provided the best-fitting modeled CMD to the data, but our overall interpretation is valid for the profiles derived from all three stellar libraries. Two of the very low-mass dSphs (CVn~II and Leo~IV) lie $2-4$ magnitudes below the rest of the sample in both initial and present-day luminosity. Excluding these two systems, the rest of the sample shows a much narrower range in their initial luminosities than their present day luminosities (i.e., $\Delta M_V\sim2$ versus $\Delta M_V\sim5$). The larger range at the present-epoch is mainly driven by changes in three galaxies. On the bright end are DDO~210 and Leo~A which show dramatic increases in star formation in the last 6 Gyr \citep{Cole2007, Cole2014}. On the faint end is Hercules which has formed far fewer stars at recent times than the initially comparable low-luminosity dIrr, Leo~T.

Looking at the early evolution of the profiles in Figure~\ref{fig:lum_time}, Leo~P's initial luminosity is similar to one very low-mass dSph (Hercules) and lower than three other dSphs (CVn~I, Draco, Ursa Minor).This similarity suggests that Leo~P may be what a dSph would look like if it evolved in isolation and retained its gas. One of the arguments against a dIrr evolving into a dSph is based on conservation of angular momentum; dIrrs are typically rotation supported whereas dSph are pressure supported \citep[e.g.,][]{Grebel2003}. However, analysis of the gas velocity field in Leo~P shows that the gas velocity dispersion approaches the rotation speed. As discussed in detail in \citet{Bernstein-Cooper2014}, there is significant random motion of the gas in Leo~P superposed on the bulk rotation signature.  This agrees with previous findings of little evidence of rotational support in other very low-mass dIrrs such as Leo~A and DDO~210 \citep{Lo1993, Young1997}. In contrast to higher-mass dIrrs with clearly defined rotation curves, the transformation of a very low-mass galaxy like Leo~P to a dSph would not require a drastic loss of angular momentum. 

\subsection{Similarities {\bf within} Morphological Types at Recent Times}

From Figure~\ref{fig:lum_time}, the patterns in luminosity profiles within the two morphological groups are quite similar over the last $\sim7$ Gyr. The dIrrs show fairly steady or increasing luminosities whereas the dSphs show declining luminosities. This more recent evolution in luminosity is presumably dictated by environment, described by the morphology-density relation (i.e., the isolated low-mass galaxies which retain their gas continue forming stars while those in closer proximity to a massive galaxy do not retain their gas and begin to fade). This is indirect evidence that the removal of gas in the dSphs is driven by environment $-$ not internal stellar feedback $-$ although feedback may be necessary to oust the gas from the inner, denser parts of the galaxies into an enveloping diffuse gaseous halo \citep[see, e.g.,][and references therein]{Mayer2006, Simpson2013, Milosavljevic2014} for more swift and efficient removal. Previous authors have attributed the removal of gas and metals in Draco, CVn~I, and Ursa Minor to stellar feedback processes \citep{Kirby2011c}, but additional factors must be involved to explain the distinct retention of gas in \textit{comparably low-mass systems} such as Leo~P and Leo~T. 

\subsection{Similarities {\bf across} Morphological Types at Early Times and Implications}
Despite the heterogeneity and general divergence of the profiles between the morphological types at more recent times, there is a similarity in the {\it first 6 Gyr} regardless of present-day morphological classification. As seen in Figure~\ref{fig:lum_time}, there is a peak in luminosity at early epochs followed by a decrease. As suggested by previous studies \citep{Cole2007, Cole2014, Brown2014}, this pattern is consistent with a quelling of star formation post-reionization. In this scenario, ionizing photons from the first epochs of star formation \citep[$z = 11.3\pm1.1$;][]{Planck2014}, may heat the outer disks of the galaxies in the mass and distance ranges probed, slowing or halting further gas accretion and causing an eventual (although sometimes temporary) decline in star formation \citep[e.g.,][]{Onorbe2015}. As noted above, the galaxies at closer distances to the hostile environment of a massive galaxy are more likely to have the hot coronal gas removed by stripping mechanisms. The more distant, isolated galaxies retain the gas heated by reionization, which is likely at extended radii in the dark matter halo. Through atomic line cooling processes over Gyr-long timescales, this hot coronal gas may eventually condense onto the disk and reignite star formation, consistent with the increases noted for the more isolated galaxies (Leo~P, Leo~A, DDO~210, Leo~T). 

The impact of reionization on an individual galaxy is likely dependent on the distance to the local source of the ionizing photons, however it is unclear what distance scales are involved \citep[e.g.,][and references therein]{Weinmann2007}. Excluding Leo~P, the distances probed by our comparison sample range from 76 kpc $-$ 1000 kpc; within this range, the early evolution of the luminosity profiles do not show differences that can be obviously tied to present-epoch distances. For example, contrary to the idea that galaxies closer to the ionizing source should be more impacted by reionization, the galaxies closest to the Milky Way (Draco, Ursa Minor, CVn~II, Leo~IV) show smaller fractional increases in luminosity between $\sim8-12$ Gyr ago; the more distant galaxies do not, suggesting additional variables (e.g., mass) must be considered. The literature SFHs used to model the luminosity evolution have similarly small uncertainties at old look back times. Thus, the luminosity profiles should be robust. The exception, of course, is Leo~P, which at the larger distance of 1.6 Mpc, limits the accuracy of the ancient SFH with the current data set. The best-fitting SFH suggests star formation was damped even at the larger distance of Leo~P. However, given the uncertainties at older times, the SFH of Leo~P does not preclude a fairly constant SFR as an alternative evolution for Leo~P. If true, this would imply the damping of star formation occurs inside of $\sim1.5$ Mpc scales for a LG-type environment. 

The impact of reionization on early star formation is also predicted to be dependent on the mass of a galaxy. Theoretical simulations fine-tune a ``filtering mass'' below which reionization universally quenches star formation. While the ``filtering mass'' generally falls within the mass range probed here \citep[e.g.,][]{Gnedin2000, Susa2004}, we do not see evidence of universal quenching, in agreement with previous results \citep[e.g.,][]{Grebel2004, Monelli2010a, Hidalgo2011, Weisz2015}. Yet, observational studies have reported something akin to this ``knife-edge'' in star formation within the temporal limitations of the CMD analysis in a few dSphs of even lower-masses \citep[i.e., $M_* \ltsimeq 10^4$ \msun][]{Brown2014}. There is still continuing debate on whether the cessation of star formation in the aforementioned systems can be solely attributed to photoevaporization from reionization \citep{Brown2014} as these galaxies are also susceptible to environmentally-driven removal of gas, and possibly stellar feedback \citep{Kirby2011c}.

Finally, we note that while Leo~P has initial luminosity properties similar to the dSphs Hercules, Draco, CVn~I, and Ursa Minor, the overall evolution of Leo~P, based on the best-fitting SFH, is most similar to Leo~T (shifted to brighter magnitudes). Leo~T lies just outside the virial radius of the Milky Way, but is thought to have been only recently accreted \citep[$t_{infall} < 1$ Gyr;][]{Rocha2012}. The similarities in the luminosity profiles support this hypothesis as tidal stripping of gaseous material in a galaxy as low-mass as Leo~T becomes important at $\sim1.5$ times the virial radius of the Milky Way, or $\sim450$ kpc \citep{Milosavljevic2014}. Thus, it is unlikely that Leo~T would have retained its gas if it had been in residence around the Milky Way for a cosmically-significant time. Previous authors have suggested that Leo~T, as well as Leo~A, fit in the general scenario outlined above for Leo~P where star formation is quelled by reionization and reignited by late phase cold gas accretion \citep{Cole2007, Ricotti2009, Weisz2012, Clementini2012}. 

\section{Conclusion\label{conclusion} }
Using $HST$ optical imaging, we have studied the stellar populations and evolutionary history of Leo~P. We refined the distance measurement to Leo~P to $1.62\pm0.15$ Mpc based on HB stars and ten RR~Lyrae candidate stars, identified from their light curves. Using a CMD-fitting technique, we have reconstructed the SFH and chemical evolution of Leo~P. The best-fitting SFH solution favors star formation at the earliest epochs, followed by a period of quiescence, and a relatively constant star formation rate at recent times.

The closest galaxies provide the most detailed constraints on the evolution of galaxies at the faint-end of the LF.  Using luminosity evolution profiles modeled from detailed SFHs, we note a similar pattern between low-mass dIrrs and comparably low-mass dSphs at early times. The luminosity profiles of both morphological types suggest a general quelling of star formation post-reionization with no obvious distance dependency, after which marked dissimilarities arise. 

There are a number of possible explanations for the quelling of star formation and subsequent evolutionary differences in this low-mass regime, but the scenario which fits the observed trend includes the following general framework: (i) heating of the outer gaseous disks of low-mass galaxies by reionization at the earliest epochs, (ii) the cessation of gas accretion onto the galaxies, which provides a governor on the total gas mass of each system, (iii) in the case of dSphs, the stripping of the hot coronal gas in the hostile halo environment of the host galaxy (e.g., the Milky Way), (iv) in the case of isolated dIrrs, the slow cooling of the ionized gas which eventually condenses and becomes available for further star formation. The similar {\it initial} luminosities of low-mass galaxies studied provides supportive evidence that the removal of gas in post-reionization epochs has a strong environmental dependence for galaxies in this mass regime. Additionally, because both isolated and non-isolated galaxies will suffer the loss of material from their shallow potential wells due to stellar feedback processes during epochs of star formation, stellar feedback alone cannot explain the present-day luminosity differences between the galaxies.

Whether or not the SF in Leo~P at early epochs was temporarily damped by re-ionizing photons remains an open question. The SFH suggests there may have been a cessation in activity from $9-13$ Gyr ago. On the other hand, within the SFH uncertainties the solution is also consistent with a relatively constant SFR scenario with no quenching. If a more constant SFR is favored, this would imply that the impact of reionization is significantly smaller outside of a $\sim1.5$ Mpc radius from a LG-type environment. Despite this ambiguity, it is clear that Leo~P contains a population of old stars and has also retained a significant gas reservoir and continues making stars at a fairly constant rate. The initial luminosity of Leo~P is similar to dSphs, suggesting that Leo~P is what a low-luminosity dSph would look like if evolved in isolation without losing its gas and its ability to form stars. 

\section{Acknowledgements}
The authors would like to thank P. Rosenfield for incorporating the latest version of the PARSEC models into {\tt MATCH}, M. Boylin-Kolchin and D. Weisz for helpful discussions, and E. Perriello at the STScI and C. Gallart and M. Monnelli at the IAC for assistance in planning the observations. Support for this work was provided by NASA through grant GO-13376 from the Space Telescope Institute, which is operated by Aura, Inc., under NASA contract NAS5-26555. J.~M.~C is supported by NSF grant AST-1211683. The ALFALFA work at Cornell is supported by NSF grants AST-0607007 and AST-1107390 to R.~G. and M.~P.~H. and by grants from the Brinson Foundation. This research made use of NASA's Astrophysical Data System and the NASA/IPAC Extragalactic Database (NED) which is operated by the Jet Propulsion Laboratory, California Institute of Technology, under contract with the National Aeronautics and Space Administration.

{\it Facilities:} \facility{Hubble Space Telescope}

\renewcommand\bibname{{References}}
\bibliography{../../bibliography.bib}

\begin{thebibliography}{}
\expandafter\ifx\csname natexlab\endcsname\relax\def\natexlab#1{#1}\fi

\bibitem[{{Asplund} {et~al.}(2009){Asplund}, {Grevesse}, {Sauval}, \&
  {Scott}}]{Asplund2009}
{Asplund}, M., {Grevesse}, N., {Sauval}, A.~J., \& {Scott}, P. 2009, \araa, 47,
  481

\bibitem[{{Babul} \& {Rees}(1992)}]{Babul1992}
{Babul}, A., \& {Rees}, M.~J. 1992, \mnras, 255, 346

\bibitem[{{Begum} {et~al.}(2008){Begum}, {Chengalur}, {Karachentsev},
  {Sharina}, \& {Kaisin}}]{Begum2008}
{Begum}, A., {Chengalur}, J.~N., {Karachentsev}, I.~D., {Sharina}, M.~E., \&
  {Kaisin}, S.~S. 2008, \mnras, 386, 1667

\bibitem[{{Bell} \& {de Jong}(2001)}]{Bell2001}
{Bell}, E.~F., \& {de Jong}, R.~S. 2001, \apj, 550, 212

\bibitem[{{Belokurov} {et~al.}(2007){Belokurov}, {Evans}, {Bell}, {Irwin},
  {Hewett}, {Koposov}, {Rockosi}, {Gilmore}, {Zucker}, {Fellhauer},
  {Wilkinson}, {Bramich}, {Vidrih}, {Rix}, {Beers}, {Schneider}, {Barentine},
  {Brewington}, {Brinkmann}, {Harvanek}, {Krzesinski}, {Long}, {Pan},
  {Snedden}, {Malanushenko}, \& {Malanushenko}}]{Belokurov2007}
{Belokurov}, V., {Evans}, N.~W., {Bell}, E.~F., {et~al.} 2007, \apjl, 657, L89

\bibitem[{{Ben{\'{\i}}tez-Llambay} {et~al.}(2014){Ben{\'{\i}}tez-Llambay},
  {Navarro}, {Abadi}, {Gottloeber}, {Yepes}, {Hoffman}, \&
  {Steinmetz}}]{Benitez-Llambay2014}
{Ben{\'{\i}}tez-Llambay}, A., {Navarro}, J.~F., {Abadi}, M.~G., {et~al.} 2014,
  ArXiv e-prints, arXiv:1405.5540

\bibitem[{{Berg} {et~al.}(2012){Berg}, {Skillman}, {Marble}, {van Zee},
  {Engelbracht}, {Lee}, {Kennicutt}, {Calzetti}, {Dale}, \&
  {Johnson}}]{Berg2012}
{Berg}, D.~A., {Skillman}, E.~D., {Marble}, A.~R., {et~al.} 2012, \apj, 754, 98

\bibitem[{{Bernard} {et~al.}(2008){Bernard}, {Gallart}, {Monelli}, {Aparicio},
  {Cassisi}, {Skillman}, {Stetson}, {Cole}, {Drozdovsky}, {Hidalgo}, {Mateo},
  \& {Tolstoy}}]{Bernard2008}
{Bernard}, E.~J., {Gallart}, C., {Monelli}, M., {et~al.} 2008, \apjl, 678, L21

\bibitem[{{Bernard} {et~al.}(2009){Bernard}, {Monelli}, {Gallart},
  {Drozdovsky}, {Stetson}, {Aparicio}, {Cassisi}, {Mayer}, {Cole}, {Hidalgo},
  {Skillman}, \& {Tolstoy}}]{Bernard2009}
{Bernard}, E.~J., {Monelli}, M., {Gallart}, C., {et~al.} 2009, \apj, 699, 1742

\bibitem[{{Bernstein-Cooper} {et~al.}(2014){Bernstein-Cooper}, {Cannon},
  {Elson}, {Warren}, {Chengular}, {Skillman}, {Adams}, {Bolatto}, {Giovanelli},
  {Haynes}, {McQuinn}, {Pardy}, {Rhode}, \& {Salzer}}]{Bernstein-Cooper2014}
{Bernstein-Cooper}, E.~Z., {Cannon}, J.~M., {Elson}, E.~C., {et~al.} 2014, \aj,
  148, 35

\bibitem[{{Binggeli}(1987)}]{Binggeli1987}
{Binggeli}, B. 1987, in Nearly Normal Galaxies. From the Planck Time to the
  Present, ed. S.~M. {Faber}, 195--206

\bibitem[{{Bono} {et~al.}(2003){Bono}, {Caputo}, {Castellani}, {Marconi},
  {Storm}, \& {Degl'Innocenti}}]{Bono2003}
{Bono}, G., {Caputo}, F., {Castellani}, V., {et~al.} 2003, \mnras, 344, 1097

\bibitem[{{Boselli} {et~al.}(2009){Boselli}, {Boissier}, {Cortese}, {Buat},
  {Hughes}, \& {Gavazzi}}]{Boselli2009}
{Boselli}, A., {Boissier}, S., {Cortese}, L., {et~al.} 2009, \apj, 706, 1527

\bibitem[{{Bothwell} {et~al.}(2009){Bothwell}, {Kennicutt}, \&
  {Lee}}]{Bothwell2009}
{Bothwell}, M.~S., {Kennicutt}, R.~C., \& {Lee}, J.~C. 2009, \mnras, 400, 154

\bibitem[{{Boyer} {et~al.}(2015{\natexlab{a}}){Boyer}, {McQuinn}, {Barmby},
  {Bonanos}, {Gehrz}, {Gordon}, {Groenewegen}, {Lagadec}, {Lennon}, {Marengo},
  {Meixner}, {Skillman}, {Sloan}, {Sonneborn}, {van Loon}, \&
  {Zijlstra}}]{Boyer2015a}
{Boyer}, M.~L., {McQuinn}, K.~B.~W., {Barmby}, P., {et~al.} 2015{\natexlab{a}},
  \apjs, 216, 10

\bibitem[{{Boyer} {et~al.}(2015{\natexlab{b}}){Boyer}, {McQuinn}, {Barmby},
  {Bonanos}, {Gehrz}, {Gordon}, {Groenewegen}, {Lagadec}, {Lennon}, {Marengo},
  {McDonald}, {Meixner}, {Skillman}, {Sloan}, {Sonneborn}, {van Loon}, \&
  {Zijlstra}}]{Boyer2015b}
---. 2015{\natexlab{b}}, \apj, 800, 51

\bibitem[{{Boylan-Kolchin} {et~al.}(2011){Boylan-Kolchin}, {Bullock}, \&
  {Kaplinghat}}]{Boylan-Kolchin2011}
{Boylan-Kolchin}, M., {Bullock}, J.~S., \& {Kaplinghat}, M. 2011, \mnras, 415,
  L40

\bibitem[{{Bressan} {et~al.}(2012){Bressan}, {Marigo}, {Girardi}, {Salasnich},
  {Dal Cero}, {Rubele}, \& {Nanni}}]{Bressan2012}
{Bressan}, A., {Marigo}, P., {Girardi}, L., {et~al.} 2012, \mnras, 427, 127

\bibitem[{{Brinchmann} {et~al.}(2004){Brinchmann}, {Charlot}, {White},
  {Tremonti}, {Kauffmann}, {Heckman}, \& {Brinkmann}}]{Brinchmann2004}
{Brinchmann}, J., {Charlot}, S., {White}, S.~D.~M., {et~al.} 2004, \mnras, 351,
  1151

\bibitem[{{Brooks} {et~al.}(2013){Brooks}, {Kuhlen}, {Zolotov}, \&
  {Hooper}}]{Brooks2013}
{Brooks}, A.~M., {Kuhlen}, M., {Zolotov}, A., \& {Hooper}, D. 2013, \apj, 765,
  22

\bibitem[{{Brown} {et~al.}(2014){Brown}, {Tumlinson}, {Geha}, {Simon},
  {Vargas}, {VandenBerg}, {Kirby}, {Kalirai}, {Avila}, {Gennaro}, {Ferguson},
  {Mu{\~n}oz}, {Guhathakurta}, \& {Renzini}}]{Brown2014}
{Brown}, T.~M., {Tumlinson}, J., {Geha}, M., {et~al.} 2014, \apj, 796, 91

\bibitem[{{Bruzual} \& {Charlot}(2003)}]{Bruzual2003}
{Bruzual}, G., \& {Charlot}, S. 2003, \mnras, 344, 1000

\bibitem[{{Cacciari}(2003)}]{Cacciari2003}
{Cacciari}, C. 2003, in Astronomical Society of the Pacific Conference Series,
  Vol. 296, New Horizons in Globular Cluster Astronomy, ed. G.~{Piotto},
  G.~{Meylan}, S.~G. {Djorgovski}, \& M.~{Riello}, 329

\bibitem[{{Cannon} {et~al.}(2011){Cannon}, {Giovanelli}, {Haynes},
  {Janowiecki}, {Parker}, {Salzer}, {Adams}, {Engstrom}, {Huang}, {McQuinn},
  {Ott}, {Saintonge}, {Skillman}, {Allan}, {Erny}, {Fliss}, \&
  {Smith}}]{Cannon2011c}
{Cannon}, J.~M., {Giovanelli}, R., {Haynes}, M.~P., {et~al.} 2011, \apjl, 739,
  L22

\bibitem[{{Carretta} {et~al.}(2000){Carretta}, {Gratton}, {Clementini}, \&
  {Fusi Pecci}}]{Carretta2000}
{Carretta}, E., {Gratton}, R.~G., {Clementini}, G., \& {Fusi Pecci}, F. 2000,
  \apj, 533, 215

\bibitem[{{Catelan}(2009)}]{Catelan2009}
{Catelan}, M. 2009, \apss, 320, 261

\bibitem[{{Chen} {et~al.}(2014){Chen}, {Girardi}, {Bressan}, {Marigo},
  {Barbieri}, \& {Kong}}]{Chen2014}
{Chen}, Y., {Girardi}, L., {Bressan}, A., {et~al.} 2014, \mnras, 444, 2525

\bibitem[{{Clementini} {et~al.}(2012){Clementini}, {Cignoni}, {Contreras
  Ramos}, {Federici}, {Ripepi}, {Marconi}, {Tosi}, \&
  {Musella}}]{Clementini2012}
{Clementini}, G., {Cignoni}, M., {Contreras Ramos}, R., {et~al.} 2012, \apj,
  756, 108

\bibitem[{{Clementini} {et~al.}(2003){Clementini}, {Gratton}, {Bragaglia},
  {Carretta}, {Di Fabrizio}, \& {Maio}}]{Clementini2003}
{Clementini}, G., {Gratton}, R., {Bragaglia}, A., {et~al.} 2003, \aj, 125, 1309

\bibitem[{{Cole} {et~al.}(2014){Cole}, {Weisz}, {Dolphin}, {Skillman},
  {McConnachie}, {Brooks}, \& {Leaman}}]{Cole2014}
{Cole}, A.~A., {Weisz}, D.~R., {Dolphin}, A.~E., {et~al.} 2014, \apj, 795, 54

\bibitem[{{Cole} {et~al.}(2007){Cole}, {Skillman}, {Tolstoy}, {Gallagher},
  {Aparicio}, {Dolphin}, {Gallart}, {Hidalgo}, {Saha}, {Stetson}, \&
  {Weisz}}]{Cole2007}
{Cole}, A.~A., {Skillman}, E.~D., {Tolstoy}, E., {et~al.} 2007, \apjl, 659, L17

\bibitem[{{C{\^o}t{\'e}} {et~al.}(2009){C{\^o}t{\'e}}, {Draginda}, {Skillman},
  \& {Miller}}]{Cote2009}
{C{\^o}t{\'e}}, S., {Draginda}, A., {Skillman}, E.~D., \& {Miller}, B.~W. 2009,
  \aj, 138, 1037

\bibitem[{{Dalcanton} {et~al.}(2009){Dalcanton}, {Williams}, {Seth}, {Dolphin},
  {Holtzman}, {Rosema}, {Skillman}, {Cole}, {Girardi}, {Gogarten},
  {Karachentsev}, {Olsen}, {Weisz}, {Christensen}, {Freeman}, {Gilbert},
  {Gallart}, {Harris}, {Hodge}, {de Jong}, {Karachentseva}, {Mateo}, {Stetson},
  {Tavarez}, {Zaritsky}, {Governato}, \& {Quinn}}]{Dalcanton2009}
{Dalcanton}, J.~J., {Williams}, B.~F., {Seth}, A.~C., {et~al.} 2009, \apjs,
  183, 67

\bibitem[{{de Jong} {et~al.}(2010){de Jong}, {Martin}, {Rix}, {Smith}, {Jin},
  \& {Macci{\`o}}}]{deJong2010}
{de Jong}, J.~T.~A., {Martin}, N.~F., {Rix}, H.-W., {et~al.} 2010, \apj, 710,
  1664

\bibitem[{{de Jong} {et~al.}(2008{\natexlab{a}}){de Jong}, {Rix}, {Martin},
  {Zucker}, {Dolphin}, {Bell}, {Belokurov}, \& {Evans}}]{deJong2008b}
{de Jong}, J.~T.~A., {Rix}, H.-W., {Martin}, N.~F., {et~al.}
  2008{\natexlab{a}}, \aj, 135, 1361

\bibitem[{{de Jong} {et~al.}(2008{\natexlab{b}}){de Jong}, {Harris}, {Coleman},
  {Martin}, {Bell}, {Rix}, {Hill}, {Skillman}, {Sand}, {Olszewski}, {Zaritsky},
  {Thompson}, {Giallongo}, {Ragazzoni}, {DiPaola}, {Farinato}, {Testa}, \&
  {Bechtold}}]{deJong2008a}
{de Jong}, J.~T.~A., {Harris}, J., {Coleman}, M.~G., {et~al.}
  2008{\natexlab{b}}, \apj, 680, 1112

\bibitem[{{Dekel} \& {Silk}(1986)}]{Dekel1986}
{Dekel}, A., \& {Silk}, J. 1986, \apj, 303, 39

\bibitem[{{Dohm-Palmer} \& {Skillman}(2002)}]{Dohm-Palmer2002}
{Dohm-Palmer}, R.~C., \& {Skillman}, E.~D. 2002, \aj, 123, 1433

\bibitem[{{Dolphin}(2000)}]{Dolphin2000}
{Dolphin}, A.~E. 2000, \pasp, 112, 1383

\bibitem[{{Dolphin}(2002)}]{Dolphin2002a}
---. 2002, \mnras, 332, 91

\bibitem[{{Dolphin}(2012)}]{Dolphin2012}
---. 2012, \apj, 751, 60

\bibitem[{{Dolphin}(2013)}]{Dolphin2013}
---. 2013, \apj, 775, 76

\bibitem[{{Dolphin} {et~al.}(2004){Dolphin}, {Saha}, {Olszewski}, {Thim},
  {Skillman}, {Gallagher}, \& {Hoessel}}]{Dolphin2004}
{Dolphin}, A.~E., {Saha}, A., {Olszewski}, E.~W., {et~al.} 2004, \aj, 127, 875

\bibitem[{{Dolphin} {et~al.}(2001){Dolphin}, {Saha}, {Skillman}, {Tolstoy},
  {Cole}, {Dohm-Palmer}, {Gallagher}, {Mateo}, \& {Hoessel}}]{Dolphin2001b}
{Dolphin}, A.~E., {Saha}, A., {Skillman}, E.~D., {et~al.} 2001, \apj, 550, 554

\bibitem[{{Dolphin} {et~al.}(2002){Dolphin}, {Saha}, {Claver}, {Skillman},
  {Cole}, {Gallagher}, {Tolstoy}, {Dohm-Palmer}, \& {Mateo}}]{Dolphin2002b}
{Dolphin}, A.~E., {Saha}, A., {Claver}, J., {et~al.} 2002, \aj, 123, 3154

\bibitem[{{Efstathiou}(1992)}]{Efstathiou1992}
{Efstathiou}, G. 1992, \mnras, 256, 43P

\bibitem[{{Einasto} {et~al.}(1974){Einasto}, {Kaasik}, \& {Saar}}]{Einasto1974}
{Einasto}, J., {Kaasik}, A., \& {Saar}, E. 1974, \nat, 250, 309

\bibitem[{{Ferrara} \& {Tolstoy}(2000)}]{Ferrara2000}
{Ferrara}, A., \& {Tolstoy}, E. 2000, \mnras, 313, 291

\bibitem[{{Ford} {et~al.}(1998){Ford}, {Bartko}, {Bely}, {Broadhurst},
  {Burrows}, {Cheng}, {Clampin}, {Crocker}, {Feldman}, {Golimowski}, {Hartig},
  {Illingworth}, {Kimble}, {Lesser}, {Miley}, {Neff}, {Postman}, {Sparks},
  {Tsvetanov}, {White}, {Sullivan}, {Krebs}, {Leviton}, {La Jeunesse},
  {Burmester}, {Fike}, {Johnson}, {Slusher}, {Volmer}, \&
  {Woodruff}}]{Ford1998}
{Ford}, H.~C., {Bartko}, F., {Bely}, P.~Y., {et~al.} 1998, in Society of
  Photo-Optical Instrumentation Engineers (SPIE) Conference Series, Vol. 3356,
  Space Telescopes and Instruments V, ed. P.~Y. {Bely} \& J.~B. {Breckinridge},
  234--248

\bibitem[{{Frebel} {et~al.}(2010){Frebel}, {Simon}, {Geha}, \&
  {Willman}}]{Frebel2010}
{Frebel}, A., {Simon}, J.~D., {Geha}, M., \& {Willman}, B. 2010, \apj, 708, 560

\bibitem[{{Gallart} {et~al.}(2005){Gallart}, {Zoccali}, \&
  {Aparicio}}]{Gallart2005}
{Gallart}, C., {Zoccali}, M., \& {Aparicio}, A. 2005, \araa, 43, 387

\bibitem[{{Giovanelli} {et~al.}(2005){Giovanelli}, {Haynes}, {Kent},
  {Perillat}, {Saintonge}, {Brosch}, {Catinella}, {Hoffman}, {Stierwalt},
  {Spekkens}, {Lerner}, {Masters}, {Momjian}, {Rosenberg}, {Springob},
  {Boselli}, {Charmandaris}, {Darling}, {Davies}, {Garcia Lambas}, {Gavazzi},
  {Giovanardi}, {Hardy}, {Hunt}, {Iovino}, {Karachentsev}, {Karachentseva},
  {Koopmann}, {Marinoni}, {Minchin}, {Muller}, {Putman}, {Pantoja}, {Salzer},
  {Scodeggio}, {Skillman}, {Solanes}, {Valotto}, {van Driel}, \& {van
  Zee}}]{Giovanelli2005}
{Giovanelli}, R., {Haynes}, M.~P., {Kent}, B.~R., {et~al.} 2005, \aj, 130, 2598

\bibitem[{{Giovanelli} {et~al.}(2013){Giovanelli}, {Haynes}, {Adams}, {Cannon},
  {Rhode}, {Salzer}, {Skillman}, {Bernstein-Cooper}, \&
  {McQuinn}}]{Giovanelli2013}
{Giovanelli}, R., {Haynes}, M.~P., {Adams}, E.~A.~K., {et~al.} 2013, \aj, 146,
  15

\bibitem[{{Girardi} {et~al.}(2010){Girardi}, {Williams}, {Gilbert},
  {Rosenfield}, {Dalcanton}, {Marigo}, {Boyer}, {Dolphin}, {Weisz},
  {Melbourne}, {Olsen}, {Seth}, \& {Skillman}}]{Girardi2010}
{Girardi}, L., {Williams}, B.~F., {Gilbert}, K.~M., {et~al.} 2010, \apj, 724,
  1030

\bibitem[{{Gnedin}(2000)}]{Gnedin2000}
{Gnedin}, N.~Y. 2000, \apj, 542, 535

\bibitem[{{Goddard} {et~al.}(2010){Goddard}, {Kennicutt}, \&
  {Ryan-Weber}}]{Goddard2010}
{Goddard}, Q.~E., {Kennicutt}, R.~C., \& {Ryan-Weber}, E.~V. 2010, \mnras, 405,
  2791

\bibitem[{{Grebel} \& {Gallagher}(2004)}]{Grebel2004}
{Grebel}, E.~K., \& {Gallagher}, III, J.~S. 2004, \apjl, 610, L89

\bibitem[{{Grebel} {et~al.}(2003){Grebel}, {Gallagher}, \&
  {Harbeck}}]{Grebel2003}
{Grebel}, E.~K., {Gallagher}, III, J.~S., \& {Harbeck}, D. 2003, \aj, 125, 1926

\bibitem[{{Haynes} {et~al.}(2011){Haynes}, {Giovanelli}, {Martin}, {Hess},
  {Saintonge}, {Adams}, {Hallenbeck}, {Hoffman}, {Huang}, {Kent}, {Koopmann},
  {Papastergis}, {Stierwalt}, {Balonek}, {Craig}, {Higdon}, {Kornreich},
  {Miller}, {O'Donoghue}, {Olowin}, {Rosenberg}, {Spekkens}, {Troischt}, \&
  {Wilcots}}]{Haynes2011}
{Haynes}, M.~P., {Giovanelli}, R., {Martin}, A.~M., {et~al.} 2011, \aj, 142,
  170

\bibitem[{{Hidalgo} {et~al.}(2011){Hidalgo}, {Aparicio}, {Skillman}, {Monelli},
  {Gallart}, {Cole}, {Dolphin}, {Weisz}, {Bernard}, {Cassisi}, {Mayer},
  {Stetson}, {Tolstoy}, \& {Ferguson}}]{Hidalgo2011}
{Hidalgo}, S.~L., {Aparicio}, A., {Skillman}, E., {et~al.} 2011, \apj, 730, 14

\bibitem[{{Huang} {et~al.}(2012){Huang}, {Haynes}, {Giovanelli}, {Brinchmann},
  {Stierwalt}, \& {Neff}}]{Huang2012}
{Huang}, S., {Haynes}, M.~P., {Giovanelli}, R., {et~al.} 2012, \aj, 143, 133

\bibitem[{{Hughes} {et~al.}(2008){Hughes}, {Wallerstein}, \&
  {Bossi}}]{Hughes2008}
{Hughes}, J., {Wallerstein}, G., \& {Bossi}, A. 2008, \aj, 136, 2321

\bibitem[{{Hunter} \& {Elmegreen}(2004)}]{Hunter2004}
{Hunter}, D.~A., \& {Elmegreen}, B.~G. 2004, \aj, 128, 2170

\bibitem[{{Hunter} {et~al.}(2012){Hunter}, {Ficut-Vicas}, {Ashley}, {Brinks},
  {Cigan}, {Elmegreen}, {Heesen}, {Herrmann}, {Johnson}, {Oh}, {Rupen},
  {Schruba}, {Simpson}, {Walter}, {Westpfahl}, {Young}, \&
  {Zhang}}]{Hunter2012}
{Hunter}, D.~A., {Ficut-Vicas}, D., {Ashley}, T., {et~al.} 2012, \aj, 144, 134

\bibitem[{{Irwin} \& {Hatzidimitriou}(1995)}]{Irwin1995}
{Irwin}, M., \& {Hatzidimitriou}, D. 1995, \mnras, 277, 1354

\bibitem[{{Kauffmann} {et~al.}(1993){Kauffmann}, {White}, \&
  {Guiderdoni}}]{Kauffmann1993}
{Kauffmann}, G., {White}, S.~D.~M., \& {Guiderdoni}, B. 1993, \mnras, 264, 201

\bibitem[{{Kennicutt} {et~al.}(1994){Kennicutt}, {Tamblyn}, \&
  {Congdon}}]{Kennicutt1994}
{Kennicutt}, Jr., R.~C., {Tamblyn}, P., \& {Congdon}, C.~E. 1994, \apj, 435, 22

\bibitem[{{Kirby} {et~al.}(2011{\natexlab{a}}){Kirby}, {Cohen}, {Smith},
  {Majewski}, {Sohn}, \& {Guhathakurta}}]{Kirby2011a}
{Kirby}, E.~N., {Cohen}, J.~G., {Smith}, G.~H., {et~al.} 2011{\natexlab{a}},
  \apj, 727, 79

\bibitem[{{Kirby} {et~al.}(2011{\natexlab{b}}){Kirby}, {Martin}, \&
  {Finlator}}]{Kirby2011c}
{Kirby}, E.~N., {Martin}, C.~L., \& {Finlator}, K. 2011{\natexlab{b}}, \apjl,
  742, L25

\bibitem[{{Kirby} {et~al.}(2010){Kirby}, {Guhathakurta}, {Simon}, {Geha},
  {Rockosi}, {Sneden}, {Cohen}, {Sohn}, {Majewski}, \& {Siegel}}]{Kirby2010}
{Kirby}, E.~N., {Guhathakurta}, P., {Simon}, J.~D., {et~al.} 2010, \apjs, 191,
  352

\bibitem[{{Kleyna} {et~al.}(2005){Kleyna}, {Wilkinson}, {Evans}, \&
  {Gilmore}}]{Kleyna2005}
{Kleyna}, J.~T., {Wilkinson}, M.~I., {Evans}, N.~W., \& {Gilmore}, G. 2005,
  \apjl, 630, L141

\bibitem[{{Klypin} {et~al.}(1999){Klypin}, {Kravtsov}, {Valenzuela}, \&
  {Prada}}]{Klypin1999}
{Klypin}, A., {Kravtsov}, A.~V., {Valenzuela}, O., \& {Prada}, F. 1999, \apj,
  522, 82

\bibitem[{{Kobulnicky} {et~al.}(2014){Kobulnicky}, {Kiminki}, {Lundquist},
  {Burke}, {Chapman}, {Keller}, {Lester}, {Rolen}, {Topel}, {Bhattacharjee},
  {Smullen}, {Vargas {\'A}lvarez}, {Runnoe}, {Dale}, \&
  {Brotherton}}]{Kobulnicky2014}
{Kobulnicky}, H.~A., {Kiminki}, D.~C., {Lundquist}, M.~J., {et~al.} 2014,
  \apjs, 213, 34

\bibitem[{{Koda} {et~al.}(2012){Koda}, {Yagi}, {Boissier}, {Gil de Paz},
  {Imanishi}, {Donovan Meyer}, {Madore}, \& {Thilker}}]{Koda2012}
{Koda}, J., {Yagi}, M., {Boissier}, S., {et~al.} 2012, \apj, 749, 20

\bibitem[{{Koposov} {et~al.}(2015){Koposov}, {Belokurov}, {Torrealba}, \&
  {Evans}}]{Koposov2015}
{Koposov}, S.~E., {Belokurov}, V., {Torrealba}, G., \& {Evans}, N.~W. 2015,
  \apj, 805, 130

\bibitem[{{Kroupa}(2001)}]{Kroupa2001}
{Kroupa}, P. 2001, \mnras, 322, 231

\bibitem[{{Lafler} \& {Kinman}(1965)}]{Lafler1965}
{Lafler}, J., \& {Kinman}, T.~D. 1965, \apjs, 11, 216

\bibitem[{{Larson}(1974)}]{Larson1974}
{Larson}, R.~B. 1974, \mnras, 169, 229

\bibitem[{{Lee} {et~al.}(2009){Lee}, {Kennicutt}, {Funes}, {Sakai}, \&
  {Akiyama}}]{Lee2009b}
{Lee}, J.~C., {Kennicutt}, Jr., R.~C., {Funes}, S.~J.~J.~G., {Sakai}, S., \&
  {Akiyama}, S. 2009, \apj, 692, 1305

\bibitem[{{Lelli} {et~al.}(2014){Lelli}, {Fraternali}, \&
  {Verheijen}}]{Lelli2014a}
{Lelli}, F., {Fraternali}, F., \& {Verheijen}, M. 2014, \aap, 563, A27

\bibitem[{{Lo} {et~al.}(1993){Lo}, {Sargent}, \& {Young}}]{Lo1993}
{Lo}, K.~Y., {Sargent}, W.~L.~W., \& {Young}, K. 1993, \aj, 106, 507

\bibitem[{{Mac Low} \& {Ferrara}(1999)}]{MacLow1999}
{Mac Low}, M., \& {Ferrara}, A. 1999, \apj, 513, 142

\bibitem[{{Makarov} {et~al.}(2006){Makarov}, {Makarova}, {Rizzi}, {Tully},
  {Dolphin}, {Sakai}, \& {Shaya}}]{Makarov2006}
{Makarov}, D., {Makarova}, L., {Rizzi}, L., {et~al.} 2006, \aj, 132, 2729

\bibitem[{{Marigo} {et~al.}(2008){Marigo}, {Girardi}, {Bressan}, {Groenewegen},
  {Silva}, \& {Granato}}]{Marigo2008}
{Marigo}, P., {Girardi}, L., {Bressan}, A., {et~al.} 2008, \aap, 482, 883

\bibitem[{{Martin} {et~al.}(2008){Martin}, {de Jong}, \& {Rix}}]{Martin2008}
{Martin}, N.~F., {de Jong}, J.~T.~A., \& {Rix}, H.-W. 2008, \apj, 684, 1075

\bibitem[{{Martin} {et~al.}(2007){Martin}, {Ibata}, {Chapman}, {Irwin}, \&
  {Lewis}}]{Martin2007}
{Martin}, N.~F., {Ibata}, R.~A., {Chapman}, S.~C., {Irwin}, M., \& {Lewis},
  G.~F. 2007, \mnras, 380, 281

\bibitem[{{Mayer} {et~al.}(2006){Mayer}, {Mastropietro}, {Wadsley}, {Stadel},
  \& {Moore}}]{Mayer2006}
{Mayer}, L., {Mastropietro}, C., {Wadsley}, J., {Stadel}, J., \& {Moore}, B.
  2006, \mnras, 369, 1021

\bibitem[{{McConnachie}(2012)}]{McConnachie2012}
{McConnachie}, A.~W. 2012, \aj, 144, 4

\bibitem[{{McQuinn} {et~al.}(2011){McQuinn}, {Skillman}, {Dalcanton},
  {Dolphin}, {Holtzman}, {Weisz}, \& {Williams}}]{McQuinn2011}
{McQuinn}, K.~B.~W., {Skillman}, E.~D., {Dalcanton}, J.~J., {et~al.} 2011,
  \apj, 740, 48

\bibitem[{{McQuinn} {et~al.}(2015{\natexlab{a}}){McQuinn}, {Skillman},
  {Dolphin}, \& {Mitchell}}]{McQuinn2015c}
{McQuinn}, K.~B.~W., {Skillman}, E.~D., {Dolphin}, A.~E., \& {Mitchell}, N.~P.
  2015{\natexlab{a}}, \apj, 808, 109

\bibitem[{{McQuinn} {et~al.}(2010){McQuinn}, {Skillman}, {Cannon}, {Dalcanton},
  {Dolphin}, {Hidalgo-Rodr{\'{\i}}guez}, {Holtzman}, {Stark}, {Weisz}, \&
  {Williams}}]{McQuinn2010a}
{McQuinn}, K.~B.~W., {Skillman}, E.~D., {Cannon}, J.~M., {et~al.} 2010, \apj,
  721, 297

\bibitem[{{McQuinn} {et~al.}(2013){McQuinn}, {Skillman}, {Berg}, {Cannon},
  {Salzer}, {Adams}, {Dolphin}, {Giovanelli}, {Haynes}, \&
  {Rhode}}]{McQuinn2013}
{McQuinn}, K.~B.~W., {Skillman}, E.~D., {Berg}, D., {et~al.} 2013, \aj, 146,
  145

\bibitem[{{McQuinn} {et~al.}(2015{\natexlab{b}}){McQuinn}, {Cannon}, {Dolphin},
  {Skillman}, {Haynes}, {Simones}, {Salzer}, {Adams}, {Elson}, {Giovanelli}, \&
  {Ott}}]{McQuinn2015a}
{McQuinn}, K.~B.~W., {Cannon}, J.~M., {Dolphin}, A.~E., {et~al.}
  2015{\natexlab{b}}, \apj, 802, 66

\bibitem[{{M{\'e}ndez} {et~al.}(2002){M{\'e}ndez}, {Davis}, {Moustakas},
  {Newman}, {Madore}, \& {Freedman}}]{Mendez2002}
{M{\'e}ndez}, B., {Davis}, M., {Moustakas}, J., {et~al.} 2002, \aj, 124, 213

\bibitem[{{Milosavljevi{\'c}} \& {Bromm}(2014)}]{Milosavljevic2014}
{Milosavljevi{\'c}}, M., \& {Bromm}, V. 2014, \mnras, 440, 50

\bibitem[{{Monelli} {et~al.}(2010{\natexlab{a}}){Monelli}, {Hidalgo},
  {Stetson}, {Aparicio}, {Gallart}, {Dolphin}, {Cole}, {Weisz}, {Skillman},
  {Bernard}, {Mayer}, {Navarro}, {Cassisi}, {Drozdovsky}, \&
  {Tolstoy}}]{Monelli2010a}
{Monelli}, M., {Hidalgo}, S.~L., {Stetson}, P.~B., {et~al.} 2010{\natexlab{a}},
  \apj, 720, 1225

\bibitem[{{Monelli} {et~al.}(2010{\natexlab{b}}){Monelli}, {Gallart},
  {Hidalgo}, {Aparicio}, {Skillman}, {Cole}, {Weisz}, {Mayer}, {Bernard},
  {Cassisi}, {Dolphin}, {Drozdovsky}, \& {Stetson}}]{Monelli2010b}
{Monelli}, M., {Gallart}, C., {Hidalgo}, S.~L., {et~al.} 2010{\natexlab{b}},
  \apj, 722, 1864

\bibitem[{{Moore} {et~al.}(1999){Moore}, {Ghigna}, {Governato}, {Lake},
  {Quinn}, {Stadel}, \& {Tozzi}}]{Moore1999}
{Moore}, B., {Ghigna}, S., {Governato}, F., {et~al.} 1999, \apjl, 524, L19

\bibitem[{{Mu{\~n}oz} {et~al.}(2006){Mu{\~n}oz}, {Carlin}, {Frinchaboy},
  {Nidever}, {Majewski}, \& {Patterson}}]{Munoz2006}
{Mu{\~n}oz}, R.~R., {Carlin}, J.~L., {Frinchaboy}, P.~M., {et~al.} 2006, \apjl,
  650, L51

\bibitem[{{Norris} {et~al.}(2010){Norris}, {Wyse}, {Gilmore}, {Yong}, {Frebel},
  {Wilkinson}, {Belokurov}, \& {Zucker}}]{Norris2010}
{Norris}, J.~E., {Wyse}, R.~F.~G., {Gilmore}, G., {et~al.} 2010, \apj, 723,
  1632

\bibitem[{{O{\~n}orbe} {et~al.}(2015){O{\~n}orbe}, {Boylan-Kolchin}, {Bullock},
  {Hopkins}, {Ker{\v e}s}, {Faucher-Gigu{\`e}re}, {Quataert}, \&
  {Murray}}]{Onorbe2015}
{O{\~n}orbe}, J., {Boylan-Kolchin}, M., {Bullock}, J.~S., {et~al.} 2015, ArXiv
  e-prints, arXiv:1502.02036

\bibitem[{{Okamoto} {et~al.}(2008){Okamoto}, {Arimoto}, {Yamada}, \&
  {Onodera}}]{Okamoto2008}
{Okamoto}, S., {Arimoto}, N., {Yamada}, Y., \& {Onodera}, M. 2008, \aap, 487,
  103

\bibitem[{{Okamoto} {et~al.}(2012){Okamoto}, {Arimoto}, {Yamada}, \&
  {Onodera}}]{Okamoto2012}
---. 2012, \apj, 744, 96

\bibitem[{{Pecaut} \& {Mamajek}(2013)}]{Pecaut2013}
{Pecaut}, M.~J., \& {Mamajek}, E.~E. 2013, \apjs, 208, 9

\bibitem[{{Pflamm-Altenburg} {et~al.}(2007){Pflamm-Altenburg}, {Weidner}, \&
  {Kroupa}}]{Pflamm-Altenburg2007}
{Pflamm-Altenburg}, J., {Weidner}, C., \& {Kroupa}, P. 2007, \apj, 671, 1550

\bibitem[{{Pietrinferni} {et~al.}(2004){Pietrinferni}, {Cassisi}, {Salaris}, \&
  {Castelli}}]{Pietrinferni2004}
{Pietrinferni}, A., {Cassisi}, S., {Salaris}, M., \& {Castelli}, F. 2004, \apj,
  612, 168

\bibitem[{{Planck Collaboration} {et~al.}(2014){Planck Collaboration}, {Ade},
  {Aghanim}, {Alves}, {Armitage-Caplan}, {Arnaud}, {Ashdown},
  {Atrio-Barandela}, {Aumont}, {Aussel}, \& et~al.}]{Planck2014}
{Planck Collaboration}, {Ade}, P.~A.~R., {Aghanim}, N., {et~al.} 2014, \aap,
  571, A1

\bibitem[{{Rhode} {et~al.}(2013){Rhode}, {Salzer}, {Haurberg}, {Van Sistine},
  {Young}, {Haynes}, {Giovanelli}, {Cannon}, {Skillman}, {McQuinn}, \&
  {Adams}}]{Rhode2013}
{Rhode}, K.~L., {Salzer}, J.~J., {Haurberg}, N.~C., {et~al.} 2013, \aj, 145,
  149

\bibitem[{{Ricotti}(2009)}]{Ricotti2009}
{Ricotti}, M. 2009, \mnras, 392, L45

\bibitem[{{Rizzi} {et~al.}(2007){Rizzi}, {Tully}, {Makarov}, {Makarova},
  {Dolphin}, {Sakai}, \& {Shaya}}]{Rizzi2007}
{Rizzi}, L., {Tully}, R.~B., {Makarov}, D., {et~al.} 2007, \apj, 661, 815

\bibitem[{{Rocha} {et~al.}(2012){Rocha}, {Peter}, \& {Bullock}}]{Rocha2012}
{Rocha}, M., {Peter}, A.~H.~G., \& {Bullock}, J. 2012, \mnras, 425, 231

\bibitem[{{Saha} {et~al.}(2011){Saha}, {Shaw}, {Claver}, \&
  {Dolphin}}]{Saha2011}
{Saha}, A., {Shaw}, R.~A., {Claver}, J.~A., \& {Dolphin}, A.~E. 2011, \pasp,
  123, 481

\bibitem[{{Sand} {et~al.}(2009){Sand}, {Olszewski}, {Willman}, {Zaritsky},
  {Seth}, {Harris}, {Piatek}, \& {Saha}}]{Sand2009}
{Sand}, D.~J., {Olszewski}, E.~W., {Willman}, B., {et~al.} 2009, \apj, 704, 898

\bibitem[{{Sand} {et~al.}(2010){Sand}, {Seth}, {Olszewski}, {Willman},
  {Zaritsky}, \& {Kallivayalil}}]{Sand2010}
{Sand}, D.~J., {Seth}, A., {Olszewski}, E.~W., {et~al.} 2010, \apj, 718, 530

\bibitem[{{Sandage}(2006)}]{Sandage2006}
{Sandage}, A. 2006, \aj, 131, 1750

\bibitem[{{Sandage} {et~al.}(1979){Sandage}, {Tammann}, \&
  {Yahil}}]{Sandage1979}
{Sandage}, A., {Tammann}, G.~A., \& {Yahil}, A. 1979, \apj, 232, 352

\bibitem[{{Sawala} {et~al.}(2015){Sawala}, {Frenk}, {Fattahi}, {Navarro},
  {Bower}, {Crain}, {Dalla Vecchia}, {Furlong}, {Jenkins}, {McCarthy}, {Qu},
  {Schaller}, {Schaye}, \& {Theuns}}]{Sawala2015}
{Sawala}, T., {Frenk}, C.~S., {Fattahi}, A., {et~al.} 2015, \mnras, 448, 2941

\bibitem[{{Schlafly} \& {Finkbeiner}(2011)}]{Schlafly2011}
{Schlafly}, E.~F., \& {Finkbeiner}, D.~P. 2011, \apj, 737, 103

\bibitem[{{Schlegel} {et~al.}(1998){Schlegel}, {Finkbeiner}, \&
  {Davis}}]{Schlegel1998}
{Schlegel}, D.~J., {Finkbeiner}, D.~P., \& {Davis}, M. 1998, \apj, 500, 525

\bibitem[{{Simon} \& {Geha}(2007)}]{Simon2007}
{Simon}, J.~D., \& {Geha}, M. 2007, \apj, 670, 313

\bibitem[{{Simpson} {et~al.}(2013){Simpson}, {Bryan}, {Johnston}, {Smith}, {Mac
  Low}, {Sharma}, \& {Tumlinson}}]{Simpson2013}
{Simpson}, C.~M., {Bryan}, G.~L., {Johnston}, K.~V., {et~al.} 2013, \mnras,
  432, 1989

\bibitem[{{Sirianni} {et~al.}(2005){Sirianni}, {Jee}, {Ben{\'{\i}}tez},
  {Blakeslee}, {Martel}, {Meurer}, {Clampin}, {De Marchi}, {Ford}, {Gilliland},
  {Hartig}, {Illingworth}, {Mack}, \& {McCann}}]{Sirianni2005}
{Sirianni}, M., {Jee}, M.~J., {Ben{\'{\i}}tez}, N., {et~al.} 2005, \pasp, 117,
  1049

\bibitem[{{Skillman} {et~al.}(1989){Skillman}, {Kennicutt}, \&
  {Hodge}}]{Skillman1989}
{Skillman}, E.~D., {Kennicutt}, R.~C., \& {Hodge}, P.~W. 1989, \apj, 347, 875

\bibitem[{{Skillman} {et~al.}(2013){Skillman}, {Salzer}, {Berg}, {Pogge},
  {Haurberg}, {Cannon}, {Aver}, {Olive}, {Giovanelli}, {Haynes}, {Adams},
  {McQuinn}, \& {Rhode}}]{Skillman2013}
{Skillman}, E.~D., {Salzer}, J.~J., {Berg}, D.~A., {et~al.} 2013, \aj, 146, 3

\bibitem[{{Skillman} {et~al.}(2014){Skillman}, {Hidalgo}, {Weisz}, {Monelli},
  {Gallart}, {Aparicio}, {Bernard}, {Boylan-Kolchin}, {Cassisi}, {Cole},
  {Dolphin}, {Ferguson}, {Mayer}, {Navarro}, {Stetson}, \&
  {Tolstoy}}]{Skillman2014}
{Skillman}, E.~D., {Hidalgo}, S.~L., {Weisz}, D.~R., {et~al.} 2014, \apj, 786,
  44

\bibitem[{{Smith}(1995)}]{Smith1995}
{Smith}, H.~A. 1995, Cambridge Astrophysics Series, 27

\bibitem[{{Susa} \& {Umemura}(2004)}]{Susa2004}
{Susa}, H., \& {Umemura}, M. 2004, \apj, 600, 1

\bibitem[{{Tang} {et~al.}(2014){Tang}, {Bressan}, {Rosenfield}, {Slemer},
  {Marigo}, {Girardi}, \& {Bianchi}}]{Tang2014}
{Tang}, J., {Bressan}, A., {Rosenfield}, P., {et~al.} 2014, \mnras, 445, 4287

\bibitem[{{Thoul} \& {Weinberg}(1996)}]{Thoul1996}
{Thoul}, A.~A., \& {Weinberg}, D.~H. 1996, \apj, 465, 608

\bibitem[{{Tully} {et~al.}(2006){Tully}, {Rizzi}, {Dolphin}, {Karachentsev},
  {Karachentseva}, {Makarov}, {Makarova}, {Sakai}, \& {Shaya}}]{Tully2006}
{Tully}, R.~B., {Rizzi}, L., {Dolphin}, A.~E., {et~al.} 2006, \aj, 132, 729

\bibitem[{{van den Bergh}(1994)}]{VandenBergh1994}
{van den Bergh}, S. 1994, \apj, 428, 617

\bibitem[{{Wegner}(2000)}]{Wegner2000}
{Wegner}, W. 2000, \mnras, 319, 771

\bibitem[{{Weidner} \& {Kroupa}(2005)}]{Weidner2005}
{Weidner}, C., \& {Kroupa}, P. 2005, \apj, 625, 754

\bibitem[{{Weinmann} {et~al.}(2007){Weinmann}, {Macci{\`o}}, {Iliev},
  {Mellema}, \& {Moore}}]{Weinmann2007}
{Weinmann}, S.~M., {Macci{\`o}}, A.~V., {Iliev}, I.~T., {Mellema}, G., \&
  {Moore}, B. 2007, \mnras, 381, 367

\bibitem[{{Weisz} {et~al.}(2014{\natexlab{a}}){Weisz}, {Dolphin}, {Skillman},
  {Holtzman}, {Gilbert}, {Dalcanton}, \& {Williams}}]{Weisz2014a}
{Weisz}, D.~R., {Dolphin}, A.~E., {Skillman}, E.~D., {et~al.}
  2014{\natexlab{a}}, \apj, 789, 147

\bibitem[{{Weisz} {et~al.}(2015){Weisz}, {Dolphin}, {Skillman}, {Holtzman},
  {Gilbert}, {Dalcanton}, \& {Williams}}]{Weisz2015}
---. 2015, ArXiv e-prints, arXiv:1503.05195

\bibitem[{{Weisz} {et~al.}(2014{\natexlab{b}}){Weisz}, {Johnson}, \&
  {Conroy}}]{Weisz2014b}
{Weisz}, D.~R., {Johnson}, B.~D., \& {Conroy}, C. 2014{\natexlab{b}}, \apjl,
  794, L3

\bibitem[{{Weisz} {et~al.}(2011){Weisz}, {Dalcanton}, {Williams}, {Gilbert},
  {Skillman}, {Seth}, {Dolphin}, {McQuinn}, {Gogarten}, {Holtzman}, {Rosema},
  {Cole}, {Karachentsev}, \& {Zaritsky}}]{Weisz2011}
{Weisz}, D.~R., {Dalcanton}, J.~J., {Williams}, B.~F., {et~al.} 2011, \apj,
  739, 5

\bibitem[{{Weisz} {et~al.}(2012){Weisz}, {Zucker}, {Dolphin}, {Martin}, {de
  Jong}, {Holtzman}, {Dalcanton}, {Gilbert}, {Williams}, {Bell}, {Belokurov},
  \& {Wyn Evans}}]{Weisz2012}
{Weisz}, D.~R., {Zucker}, D.~B., {Dolphin}, A.~E., {et~al.} 2012, \apj, 748, 88

\bibitem[{{Young} \& {Lo}(1997)}]{Young1997}
{Young}, L.~M., \& {Lo}, K.~Y. 1997, \apj, 490, 710

\end{thebibliography}

\end{document}